\documentclass[12pt]{iopart}

\usepackage{iopams}
\usepackage{setstack}
\usepackage{graphicx}
\usepackage{amssymb}
\usepackage{hyperref}
\hypersetup{
 pdfnewwindow=true,
 colorlinks=true,
 linkcolor=blue,
 citecolor=blue,
 filecolor=blue,
 urlcolor=blue
}
\usepackage{color}
\usepackage{psfrag}
\usepackage{soul}
\usepackage{textcomp}
\usepackage[top=2.5cm, left=2.3cm, right=2.3cm, bottom=2cm]{geometry}

\newcommand{\eqref}[1]{(\ref{#1})}
\newcommand{\comments}[1]{}
\newcommand{\bea}{\begin{eqnarray}}
\newcommand{\eea}{\end{eqnarray}}
\newcommand{\ket}[1]{| #1 \rangle}

\newcommand{\bra}[1]{\langle #1 |}

\begin{document}

\title[Quantum current in dissipative systems]{Quantum current in dissipative systems}

\author{Karen V. Hovhannisyan and Alberto Imparato}

\address{Department of Physics and Astronomy, Aarhus University, Ny Munkegade 120, 8000 Aarhus, Denmark}
\ead{karen@phys.au.dk and imparato@phys.au.dk}

\begin{abstract}
Describing current in open quantum systems can be problematic due to the subtle interplay of quantum coherence and environmental noise. Probing the noise-induced current can be detrimental to the tunneling-induced current and vice versa. We derive a general theory for the probability current in quantum systems arbitrarily interacting with their environment that overcomes this difficulty. We show that the current can be experimentally measured by performing a sequence of weak and standard quantum measurements. We exemplify our theory by analyzing a simple Smoluchowski-Feynman-type ratchet consisting of two particles, operating deep in the quantum regime. Fully incorporating both thermal and quantum effects, the current generated in the model can be used to detect the onset of ``genuine quantumness'' in the form of quantum contextuality. The model can also be used to generate steady-state entanglement in the presence of arbitrarily hot environment.
\end{abstract}

\section{Introduction}

Current is a central notion in transport theory and non-equilibrium thermodynamics. Defining quantum current in closed systems is straightforward \cite{Messiah_v1}. Yet, despite the considerable attention quantum transport and quantum walks in dissipative systems have received \cite{Gebauer_2004, Kohler_2005, Bodor_2006, Zhang_2008, Mohseni_2008, Whitfield_2010, Mulken_2011, Avron_2012, Zimboras_2013}, there exists no general definition of quantum current in open quantum systems, which would take into account both tunneling and environment-induced hopping. The issue has been addressed only in the specific case of the position operator current in continuous-variable systems with local, only position-dependent system-environment interaction, in the Markovian approximation \cite{Gebauer_2004, Bodor_2006, Avron_2012}. Moreover, unlike heat to work conversion, which is fairly well understood both classically \cite{Feynman, Seifert_2012} and quantumly \cite{Goold_2016, Benenti_2017}, motion generation solely utilizing thermal disequilibrium has mainly remained in the focus of classical thermodynamics and transport \cite{Feynman, Magnasco_1993, Reimann_2002, Gomez-Marin_2005, Hanggi_2009, Seifert_2012}. Only a handful of models have been studied in the quantum regime: Except for several works \cite{Mari_2015, Roulet_2017, Seah_2018, Roulet_2018, Fogedby_2018} studying the (angular) momentum in continuous-variable thermal rotors, to our knowledge, only Ref.~\cite{Bissbort_2017} studies persistent current generation in a fully quantum model, albeit in the limit of the moving part being isolated from the environment. 

In any realistic model, however, all parts of the device will, in one way or another, be coupled to the environment (think of all models inherited from the classical domain, e.g., the Smoluchowski-Feynman ratchet \cite{Smoluchowski_1912, Feynman}). Therefore, in order to study current generation in quantum devices, one first has to deal with the fundamental problem of defining the current in open quantum systems. We solve this problem in the most general form, by deriving a surprisingly simple formula for probability current, universally applicable to systems undergoing arbitrary dynamics (be it Markovian or non-Markovian). We furthermore establish a link between the current and quantum weak measurements, which allows us to gain valuable intuition about the theory.

We illustrate the power of our theory on a minimalist model of a quantum rotor, capable of autonomously generating current in the steady-state regime. The model consists of two particles with $3$-dimensional Hilbert spaces, interacting with each other through a quantum Potts Hamiltonian \cite{Wu_1982}, which is essentially a ``higher-spin'' (spin-$1$, in our case) generalization of the Ising model, complemented with quantum tunneling. Each particle is coupled to its own thermal environment, making the model an ideal testbed for our theory (no part is isolated). Alternatively, one can think of two atoms on an optical lattice \cite{Greiner_2008}, each of which is confined to $3$ lattice sites. We restrict to the $3$-position model as it is the minimal case exhibiting the symmetry breaking necessary for ordered motion to occur. Using standard techniques from the theory of open quantum systems, we fully characterize the non-equilibrium steady states of the rotor, allowing us to make a rigorous connection between the symmetries of the rotor's Hamiltonian and the main transport properties of the system: particle current and heat flux. Despite its simplicity, the model exhibits several non-trivial quantum effects which have never been reported in the literature before. For example, the following highly counterintuitive current inversion phenomenon takes place: Although tunneling does not favour any specific direction of particle propagation, the particle current can change its direction as the rate of tunneling is varied. Moreover, it turns out that, in our model, this effect is always preceded by the onset of genuine (classically unsimulatable) quantumness, expressed via quantum contextuality, in the system. Lastly, enabled by global rotation symmetry, our machine is capable of converting locally coherent, yet completely uncorrelated states of the rotor into entangled steady states, for \textit{arbitrarily high} temperatures of the baths. Moreover, in the presence of sufficiently strong tunneling, powered by temperature difference, the machine can ``charge'' the initially ``empty'' (i.e., incapable of producing work) rotor with extractable work.

\section{Defining Current}
\label{sec:current}

Imagine a classical stochastic process realized by a particle hopping on a graph with vertices $\{ j, j', ... \}$; this is the most general representation of any part of any classical thermal machine. The probability current---which defines the mass/charge current and information transport---between vertices $j$ and $j'$, at any moment of time, is quite standardly defined as \cite{Seifert_2012}
\bea \label{eq:clacur}
J_{j\to j'} = p_j W_{j|j'} - p_{j'} W_{j'|j},
\eea
where $p_j$ is the probability of finding the particle at vertex $j$ and $W_{j|j'}$ is the transition rate from vertex $j$ to $j'$. 

Surprisingly, no general quantum analogue of Eq.~\eqref{eq:clacur} exists, even when the particle's open-system dynamics can be described by a standard Gorini-Kossakowski-Lindblad-Sudarshan (GKLS) master equation (ME) \cite{bp} (e.g., the Born-Markov and secular approximations are applicable \cite{bp}):
\bea \label{eq:GKLS}
\frac{d \rho}{d t} = - \mathrm{i} [H, \rho] + \sum_{\lambda} \gamma_\lambda \mathcal{S}[\Lambda_\lambda][\rho].
\eea
Here, $\rho$ is the state of the particle, $H$ is its Hamiltonian, and $\mathcal{S}[\Lambda][\rho] = \Lambda \rho \Lambda^\dagger - \frac{1}{2} \{ \Lambda^\dagger \Lambda, \rho \}$, with $\{ \cdot, \cdot \}$ denoting the anticommutator. To gain some intuition, let us for a moment assume that being at each vertex corresponds to a pure state, and that these states constitute an orthonormal basis of the particle's Hilbert space (i.e., that the particle has no internal structure). Then, the classical picture, including the current being given by Eq.~\eqref{eq:clacur}, is recovered in the trivial case when $\rho$ and $H$ are diagonal in the vertex basis at all times and the ``jump'' operators, $\Lambda_\lambda$, are of the form $\ket{j}\bra{j'}$. This is due to the fact that, in such cases, the quantum evolution, determined entirely by the dynamics of $\rho$'s diagonal elements, exactly coincides with the classical stochastic dynamics.

However, whenever the Hamiltonian is not diagonal in the vertex basis (in other words, if tunneling is possible between vertices), defining the current becomes problematic. Indeed, although the probabilities to find the particle at given vertices are always well defined, defining transition rates between vertices is less straightforward: The classical protocol---measure at $j$, evolve, measure at $j'$---executed in the quantum regime \cite{Leggett_1984, Mohseni_2008, Whitfield_2010, Zimboras_2013}, excludes the coherent part of the dynamics, and hence cannot provide a complete picture.

In order to find a proper definition for the current, we will undertake a more consistent approach---standardly used in quantum transport theory---to defining current, which is based on the continuity equation. In this approach, the current of an operator $x$, $J$, is inferred from the fact that, in the Heisenberg picture, the continuity equation $d x/dt = - \mathrm{div} J$ must hold \cite{Caroli_1971, Camalet_2003, Gebauer_2004, Kohler_2005, Bodor_2006} (note that current is not to be confused with velocity \cite{Messiah_v1, Albarelli_2016}). When the configuration space is discrete, as is the case here, the continuity equation takes the form
\bea \label{eq:div3}
\frac{d x_j}{dt} = \sum_{j' \neq j} J_{j' \to j},
\eea
where the subscript $j$ means that the operator $x$ is taken at the position (vertex) $j$. Since we are interested in the particle current, from now on, the operator $x_j$, in the Schr\"{o}dinger picture, will be the projector on the $j$'th vertex (for example, in the above simplified picture where ``the particle is at vertex $j$'' was equivalent to ``the particle is in pure state $\ket{j}$'', one would have $x_j = \ket{j}\bra{j}$). Moreover, in order to determine the current at the moment of time $t$, the transition from the Schr\"{o}dinger to Heisenberg picture has to be performed at the very same moment of time $t$. Such a reservation has to be made in order to ensure that the current is actually being calculated at a given location (see \ref{app:Heis}).

Now, keeping in mind that $\sum_j x_j = \mathbb{I}$ (where $\mathbb{I}$ is the identity operator on the full Hilbert space of the particle) holds at all times in the Heisenberg picture, it is straightforward to show that the quantities
\bea \label{eq:current}
J_{j \to j'} = \frac{1}{2} \left\{ x_j,  \frac{d {x}_{j'}}{d t} \right\} - \frac{1}{2} \left\{ x_{j'}, \frac{d {x}_{j}}{d t}\right\}
\eea
satisfy the continuity equation \eqref{eq:div3}. In view of the above, note that $J$ can depend on time only through the time derivatives of $x_j$ and $x_{j'}$, and the average current is given by $\langle J_{j \to j'} \rangle (t) = \tr[\rho(t) J_{j \to j'}(t)]$. We note that Eq.~\eqref{eq:current} applies whenever $\{ x_j \}$ is a complete set of orthogonal projectors. Eq.~\eqref{eq:current} is the main result of this section: it defines the probability current in arbitrary quantum systems undergoing arbitrary dynamics, as long as it is positive and trace-preserving (see also \ref{app:Heis}).

When the evolution is governed by a GKLS equation, Eq.~\eqref{eq:GKLS}, the Heisenberg-picture operators evolve according to $\frac{d x_j}{d t} = \mathrm{i}[H, x_j] + \sum_\lambda \gamma_\lambda \mathcal{S}^{*}[\Lambda_\lambda][x_j]$, where $S^{*}$ is the dual of $S$: $S^{*}[\Lambda][x] = \Lambda^\dagger x \Lambda - \frac{1}{2} \{ \Lambda^\dagger \Lambda, x \}$. Taking into account that $x_j x_{j'} = x_j \delta_{j, j'}$, for an arbitrary GKLS ME, we obtain that $J_{j \to j'} = J_{j \to j'}^{(\mathrm{tun})} + J_{j \to j'}^{(\mathrm{th})}$, where
\bea \label{eq:tuncur}
J_{j \to j'}^{(\mathrm{tun})} = \mathrm{i} (x_{j} H x_{j'} - x_{j'} H x_{j})
\eea
is the current resulting from the vertex-to-vertex transitions caused by the internal dynamics, which is often referred to as tunneling (hence the superscript of the operator). This operator is routinely used to describe particle current in condensed-matter physics \cite{Messiah_v1, Caroli_1971, Gebauer_2004, Kohler_2005, Bodor_2006, Zhang_2008}. Note that $J_{j \to j'}^{(\mathrm{tun})}$ is essentially the discretized version of the textbook current associated with the continuous-space Schr\"{o}dinger equation \cite{Messiah_v1}: $J(q) = \frac{\hbar}{2m\mathrm{i}}\left[ \Psi(q)^* \nabla \Psi(q) - \Psi(q)\nabla \Psi^*(q) \right]$, where $\Psi(q)$ is the state in the position ($q$) representation, and the corresponding current operator is $J^{(\mathrm{Sch})}_q = \frac{1}{2m}\{\hat{p}, \ket{q}\bra{q} \}$, with $\hat{p}$ being the momentum operator (cf. Eq.~\eqref{eq:current}). In fact, in \ref{app:conti}, we show that, when one discretizes a continuous-space Hamiltonian $\frac{\hat{p}^2}{2m}+V(\hat{q})$ (where $V(\hat{q})$ is some potential), then calculates $J_{j \to j'}^{(\mathrm{tun})}$ according to Eq.~\eqref{eq:tuncur}, and then takes the continuous limit, one ends up with $J^{(\mathrm{Sch})}_q$.

The second component of the current, $J_{j \to j'}^{(\mathrm{th})}$, is the current resulting from vertex-to-vertex jumps originating from the interaction with the environment, and is given by
\bea \label{eq:thcur}
J_{j \to j'}^{(\mathrm{th})} = \frac{1}{2}\sum_\lambda \gamma_\lambda [ \{ x_{j}, \Lambda_\lambda^\dagger x_{j'} \Lambda_\lambda \} - \{ x_{j'}, \Lambda_\lambda^\dagger x_{j} \Lambda_\lambda \} ],
\eea
where, in both $J_{j \to j'}^{(\mathrm{th})}$ and $J_{j \to j'}^{(\mathrm{tun})}$, all operators are in the Schr\"{o}dinger picture (see \ref{app:Heis}). Note that Eqs.~\eqref{eq:tuncur} and \eqref{eq:thcur} apply to \textit{any} GKLS ME. An analogue of Eq.~\eqref{eq:thcur} was derived in Refs.~\cite{Gebauer_2004, Bodor_2006, Avron_2012} for the special case of continuous-variable systems which couple to their environment through local, position-dependent interaction terms (i.e., when the operator, the current of which is to be determined, commutes with the system-environment interaction Hamiltonian). We also note that, in general, the separation into unitary and dissipative parts in the Lindbladian \eqref{eq:GKLS}---and hence the division between tunneling and dissipative currents---is not unique \cite{bp}. However, in those situations where the system is coupled to the environment weakly and its Hamiltonian is known, which is the case in the example we construct in Sec.~\ref{sec:model}, the system Hamiltonian (generally renormalized by the Lamb shift Hamiltonian) represents the most natural choice for the generator of the unitary part of the GKLS equation, in the sense that eliminating the environment will nullify the dissipative current and introduce only a second-order correction to the tunneling current originating from the Lamb shift term. Lastly, let us note that the generalization of Eqs.~\eqref{eq:tuncur} and \eqref{eq:thcur} to an arbitrary time-local Redfield equation \cite{bp} is straightforward (see also \ref{app:Heis}).

As any other quantity inferred from its divergence, the current defined by Eq.~\eqref{eq:current} can be supplemented by any zero-divergence term, and it will still satisfy the continuity equation \eqref{eq:div3}. Among this infinite family of possible currents, we choose the one given by Eq.~\eqref{eq:current} for two fundamental reasons. First, in \ref{app:triangle}, for the minimal non-trivial situation---a particle on a three-vertex graph, evolving according to a GKLS ME---we show by an explicit calculation of the derivative of $x_j$ that there arise no terms other than those in Eqs.~\eqref{eq:tuncur} and \eqref{eq:thcur}. Second, as we show below, its average can be rewritten in the form of Eq.~\eqref{eq:clacur}, and, in the classical limit discussed below Eq.~\eqref{eq:GKLS}, it converges to the classical current.

The remarkably simple Eq.~\eqref{eq:current} for the current of a particle undergoing an arbitrarily general quantum evolution on an arbitrary graph, to the best of our knowledge, has not been reported in the literature before. In order to better understand its meaning, let us rewrite the derivatives in Eq.~\eqref{eq:current} as limit processes, so that the average current at the moment of time $t$ reads
\bea \label{eq:avweakcur}
\langle J_{j \to j'} \rangle (t) = \frac{\mathcal{P}_t \left[ x_{j} | \, x_{j'} (\epsilon) \right] p_{j} (t + \epsilon) - (j \leftrightarrow j')}{\epsilon} ,
\eea
where $\epsilon>0$ is much smaller than the time-scale of the system evolution (but much larger than the relaxation time-scale of the environment) and the symbol $(j \leftrightarrow j')$ denotes the repeating of the same term as on its left, but with the $j$ and $j'$ indices interchanged. Here, $p_{j'}(t + \epsilon) = \tr [\rho(t) x_{j'}(\epsilon)] = \tr [\rho(t+\epsilon) x_{j'}]$ is the probability of detecting the particle at the vertex $j'$, at the moment of time $t + \epsilon$, and
\bea \label{eq:weakval}
\mathcal{P}_t \left[ x_{j} | \, x_{j'} (\epsilon) \right] = \frac{\tr\left[ x_{j} \{ \rho (t), x_{j'} (\epsilon) \} \right]}{2 \tr \left[ x_{j'} (\epsilon) \rho (t) \right]}
\eea
is the real part of the weak value of $x_{j}(t)$ on the preselected state $\rho(t)$ and postselected on $x_{j'}(\epsilon)$ \cite{Aharonov_1988, Dressel_2010, Dressel_2014}. More specifically, it can be interpreted as the conditional average of $x_{j}$, at the moment of time $t$, conditioned on the measurement of $x_{j'}$ at a later moment $t + \epsilon$ \cite{Dressel_2010}. Since $x_{j}$ is the projector on the location $j$, the conditional average is the same as the conditional probability. In other words, $\mathcal{P}_t \left[ x_{j} | \, x_{j'} (\epsilon) \right]$ is the probability of finding the particle at the vertex $j$, at the moment of time $t$, as a result of a weak (minimal-disturbance) measurement of $x_{j}$ on the system in the state $\rho(t)$, conditioned on a measurement of the particle at the site $j'$ at a later moment of time $t + \epsilon$. In turn, this prompts the interpretation of $\mathcal{P}_t \left[ x_{j} | \, x_{j'} (\epsilon) \right] p_{j'} (t + \epsilon)$ as the joint probability of finding the particle at location $x_{j}$, at the moment of time $t$, and at location $x_{j'}$, at the moment of time $t + \epsilon$. Therefore, the formula for the current, Eq.~\eqref{eq:avweakcur}, fits the classical intuition of ``flow forward minus flow backward'', encapsulated in Eq.~\eqref{eq:clacur} (with $\lim_{\epsilon \to 0}\mathcal{P}_t/\epsilon$ playing the role of the transition rate). Importantly, the location at $t$ is measured weakly so that the state is disturbed minimally before the second measurement. In this way, the quantum coherence effects are not neglected while we observe the particle jump from vertex $j$ to $j'$. In order to appreciate the importance of the first measurement being weak, in \ref{app:TSM}, we calculate the current of a particle obeying a GKLS equation, when both position measurements are non-weak. We find that, indeed, in such a case, the coherent terms of the current, among which there is also the tunneling current, are not present anymore. Interestingly, when the vertices correspond to pure states and the jump operators are of the form $\ket{j}\bra{j'}$, the thermal current, given by Eq.~\eqref{eq:thcur}, coincides with that defined through two strong measurements (cf. Eq.~\eqref{eq:TSMcur} in \ref{app:TSM}), which was to be expected, as, with such jump operators, the environment is constantly measuring the system \cite{bp}. In this case, the average thermal current is given by $\langle J^{(\mathrm{th})}_{j \to j'} \rangle = \gamma_{j'j}\rho_{jj} - \gamma_{jj'}\rho_{j'j'}$, which makes it evident that, in this generic classical limit, the current Eq.~\eqref{eq:current} converges to the classical current Eq.~\eqref{eq:clacur}.

Note that the distribution $\mathcal{P}_t \left[ x_{j} | \, x_{j'} (\epsilon) \right] p_{j'} (t + \epsilon) = \mathrm{Re} \, \tr\left[ x_{j} x_{j'}(\epsilon) \rho(t) \right]$ is also known as Terletsky-Margenau-Hill quasiprobability distribution \cite{Terletsky_1937, Margenau_1961} (see also Ref.~\cite{Dressel_2015} for further discussion on the connection between weak value and Terletsky-Margenau-Hill distribution). We also note that, besides providing an appropriate theoretical language for interpreting Eq.~\eqref{eq:current}, weak values can also be directly accessed experimentally \cite{Dressel_2014}. This means that the average current in Eq.~\eqref{eq:avweakcur} can be measured in an experiment by literally first measuring the partcile's position weakly at site $j$ and then, after a short period of time, strongly at site $j'$. Despite the widespread applications of weak measurements in various areas of quantum physics (see, e.g., Refs.~\cite{Dressel_2014, Lostaglio_2018}), to the best of our knowledge, the above connection with quantum transport has not been previously made in the literature, even on the level of the well-known and widely used tunneling current.

What is more, Eqs.~\eqref{eq:avweakcur} and \eqref{eq:weakval}, through the results in Refs.~\cite{Pusey_2014, Lostaglio_2018}, provide us with a link between the average current and a fundamental notion of non-classicality, called contextuality \cite{Peres, Spekkens_2005}. In contrast to Bell nonlocality \cite{Peres, mikeike}, which only applies to situations involving space-like separated parties, contextuality scenarios are designed to decide whether a localized (quantum) system experiment can be simulated by a classical, deterministic hidden variable model \cite{Peres, Spekkens_2005} (see \ref{app:context} for precise definitions of these concepts). On the other hand, similarly to Bell nonlocality and entanglement, contextuality has been recently identified as a key resource in quantum-enhanced technology \cite{Howard_2014}. Now, as is shown in Refs.~\cite{Pusey_2014, Lostaglio_2018}, when the weak value, given by Eq.~\eqref{eq:weakval}, is negative, it means that there exists a generalized quantum measurement, call it $\{ V_\mu \}$, such that the state of the particle, $\rho(t)$, the measurement $\{ V_\mu \}$, and postselection $x_{j'}(\epsilon)$ cannot be simulated by any measurement-noncontextual hidden-variable model satisfying outcome determinism (\ref{app:context}). Remarkably, for the model ratchet introduced below, we will show that the onset of contextuality can be detected by observing the inversion of the thermal component of the average current (see Fig.~\ref{fig:fig2}).

\section{Generating Current} \label{sec:model}

The graph has to have at least three vertices in order for the particle current to be meaningful. Therefore, the minimal system that can harbour current is a three-dimensional Hilbert space, and the vertices of the graph correspond to three mutually orthogonal pure state which constitute a basis in that Hilbert space. In other words, the minimal system is a qutrit, which can be realized by physical objects as diverse as a spin-$1$ particle, viewed from the perspective of the eigenstates of the angular momentum in some direction, or an atom on an optical lattice \cite{Greiner_2008}, confined to hop and tunnel (the confining potential cannot be infinite) between three neighbouring potential wells.

\begin{figure}[t!] 
\centering
\includegraphics[width=0.65\textwidth]{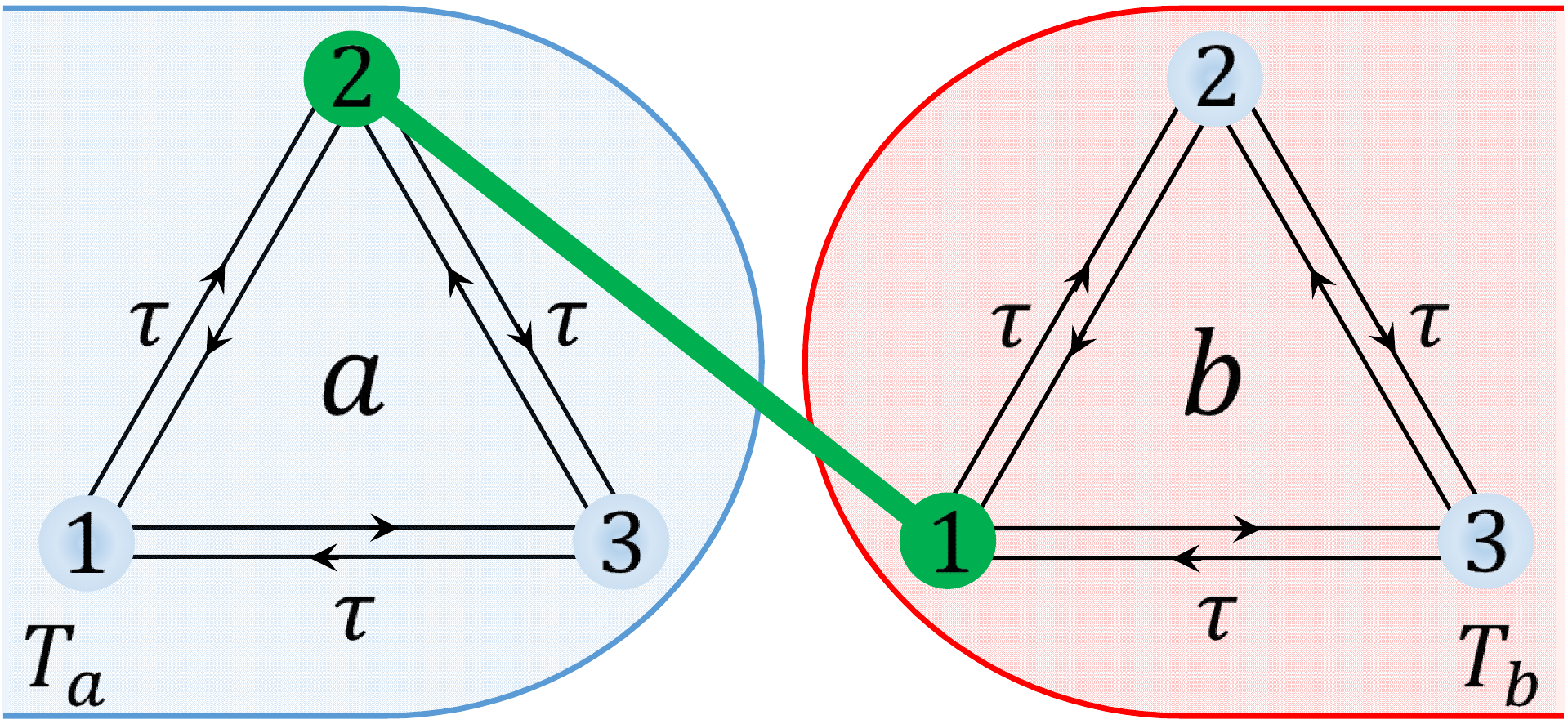}
    \caption{Schematic picture of the machine. The rotor consists of two partitions $a$ and $b$, each consisting of three locations a particle (represented by a green circle) can occupy. The particles can tunnel between sites at rate $\tau$. The interaction between the particles is illustrated by the connecting green line, and the blue and red semicircles illustrate the fact that each partition is coupled to its own thermal bath.}
\label{fig:fig1}
\end{figure}

Our model rotor consists of two such minimal systems, particle $a$ and particle $b$, each interacting with their own thermal baths, respectively, at temperatures $T_a$ and $T_b$ (see Fig.~\ref{fig:fig1} for an illustration). The self-Hamiltonians of these particles are taken to be translation (rotation) invariant and periodic, so that the ``ratchetness'', i.e., the spatial asymmetry indispensable \cite{Feynman, Magnasco_1993, Reimann_2002, Gomez-Marin_2005} for triggering the motion, is delegated to the interaction between the particles. Specifically,
\bea \label{eq:Haml}
H_{ab} = \tau X_a\otimes \mathbb{I}_b + \tau \mathbb{I}_a\otimes X_b + H_\mathrm{int},
\eea
where $\mathbb{I}$ is the identity operator in the corresponding Hilbert space and, in the basis of vertex states,
\bea \label{eq:X}
X = \sum_{k=1}^3( \ket{k+1}\bra{k} + \ket{k}\bra{k+1}).
\eea
The quantity $\tau X$ determines the local rotation-symmetric self-Hamiltonians of the particles, with $\tau$ being the tunneling rate between vertices. Here, the index is cyclic in the sense that $\ket{k+3}\equiv\ket{k}$. We choose the interaction term to be classical, $\bra{j_a, j_b} H_\mathrm{int} \ket{k_a, k_b} = \delta_{j_a k_a} \delta_{j_b k_b} U(j_a, j_b)$, with the potential $U$ being of Potts form \cite{Wu_1982}:
\bea \label{eq:poten}
U_{j_a,j_b} = \frac{K}{2} \cos\left[\frac{2\pi}{3}(j_a-j_b) + \phi \right].
\eea
Such a potential mimics a dipole-dipole interaction between two spins, each pointing to one of the three equispaced directions specified by the angles $2\pi j/3$. The potential is periodic in each coordinate, $U(j_a+3,j_b) = U(j_a,j_b+3) = U(j_a,j_b)$. However, unless $\phi=0$, it breaks the exchange symmetry in that swapping the particles bears an energetic cost: $U_{j_a,j_b} \neq U_{j_b,j_a}$. In what follows, we will see that it is this exchange symmetry breaking that is responsible for the occurrence of thermal current in our model.

Reading from Eq.~\eqref{eq:tuncur}, we obtain that the tunneling current of, say, the particle $a$ is given by
\bea \label{eq:MODELtuncur}
J^{(\mathrm{tun})}_{j_a \to j_a + 1} = \mathrm{i}\tau\left( \ket{j_a}\bra{j_a + 1} - \ket{j_a + 1}\bra{j_a} \right) \otimes \mathbb{I}_b . ~~
\eea

We will consider only the limit of weak coupling to a Markovian environment, so that the evolution of the rotor is governed by a standard GKLS ME. One can use two approaches towards prescribing a GKLS equation to our model: phenomenological and that based on microscopic derivation. Phenomenologically, one wishes a GKLS ME which, in the classical limit (expressed by $\tau \to 0$), would coincide with the classical ME: $\frac{d p_{j_a j_b}}{dt} =\sum'_{j_a',j_b'}[p_{j_a'j_b'} W_{j_a' j_b'|j_a j_b} - p_{j_a j_b} W_{j_a j_b | j_a' j_b'}]$. There, the sum is primed to indicate that it does not include simultaneous jumps, and the transition rates satisfy local detailed balance (e.g., for the jumps of particle $a$, $W_{j_a j_b | j_a' j_b}/W_{j_a' j_b | j_a j_b} = \exp\left[ (U_{j_a' j_b} - U_{j_a j_b}) / T_a \right]$). Now, it is straightforward to see that the GKLS equation \eqref{eq:GKLS} with the jump operators of the form $\ket{j_a}\bra{j_a'}\otimes \ket{j_b}\bra{j_b'}$ and the dissipation rates $\gamma_\lambda$ coinciding with the transition rates $W_{j_a' j_b' | j_a j_b}$, is equivalent to the classical ME. Thus, using these jump operators also when $\tau \neq 0$, one obtains a quantum GKLS master equation with a well-defined and correct classical limit. This is the standard quantum ME used, e.g., in the theory of quantum random walks \cite{Whitfield_2010, Mulken_2011}. This ME falls under the category of ``local'' GKLS MEs since its dissipators are of product form, and hence it cannot generally grasp the global dynamics of the system. This is typically expressed by the thermal state not being a steady-state solution of the local ME when the temperatures of the baths are equal. This brings about thermodynamically inconsistent behavior \cite{Walls_1970, Novotny_2002, Levy_2014, Gonzalez_2017, Hofer_2017_NJP} such as non-zero particle or heat flow between two thermal baths at equal temperatures \cite{Novotny_2002}, or, when the temperatures are different, spontaneous heat flow against the temperature gradient \cite{Levy_2014}.

This is contrasted by the microscopic approach, where one derives the equation for the state of the system from the exact dynamics of the system-plus-environment composite, under the Born-Markov and secular approximations \cite{bp}. The GKLS equation thus obtained takes full account of the global structure of the Hamiltonian even if the environments act locally on the subsystems, and it does not suffer from thermodynamic inconsistencies. For a general total Hamiltonian of the form $H_\mathrm{tot} = H_{ab} + \! \sum_{\alpha = a, b} A_\alpha \otimes B_\alpha + \! \sum_{\alpha = a, b} h_\alpha$, where $h_a$ and $h_b$ are the Hamiltonians of the baths and $A_\alpha$ and $B_\alpha$ are operators belonging, respectively, to the particles and the baths, the microscopically derived GKLS ME takes the following form:
\bea \label{eq:GKLSglob}
\frac{d \rho}{d t} = - \mathrm{i} [H, \rho] + \! \sum_{\omega \atop \alpha = a, b} \! \gamma_\alpha (\omega) \mathcal{S} [\Lambda_\alpha (\omega)] [\rho] \equiv \mathcal{L}_\mathrm{gl}[\rho], ~~~
\eea
with the jump operators given by
\bea
\label{eq:jumpglob}
\Lambda_{\alpha} (\omega) = \sum_{E_k - E_m = \omega} \Pi_{E_m} A_\alpha \Pi_{E_k}.
\eea
Here, $E_m$ is an eigenvalue of $H_{ab}$ and $\Pi_{E_m}$ is the eigenprojector corresponding to it. Note that the $\omega$'s are all the gaps of the system's bare Hamiltonian $H_{ab}$. Furthermore, since we are working in the weak coupling regime, we will henceforth omit the Lamb shift correction to the Hamiltonian, in view of its being a second-order effect \cite{bp}.

The bath dissipation rates are given by bath correlation functions: $\gamma_\alpha(\omega) = \int_{-\infty}^\infty ds e^{\mathrm{i}\omega s} \langle B_\alpha^\dagger(s)B_\alpha (0) \rangle$. For a generic bosonic bath, living in a single spatial dimension, we find \cite{bp}
\bea \label{eq:trrate3}
\gamma_\alpha(\omega) = \frac{g |\omega|}{1 - e^{-\beta_\alpha |\omega|}} \times \left\{ \begin{array}{cll}
e^{\beta_\alpha \omega} & \mathrm{when} & \omega \leq 0 \\
1 & \mathrm{when} & \omega > 0
\end{array} \right. . ~~~
\eea
These functions satisfy the detailed balance condition: $\frac{\gamma_\alpha(-\omega)}{\gamma_\alpha(\omega)}=e^{-\beta_\alpha \omega}$, which, combined with Eq.~\eqref{eq:jumpglob}, guarantees that, when $\beta_a = \beta_b$, the thermal state is a steady solution for the global ME given by Eq.~\eqref{eq:GKLSglob} (see Ref.~\cite{bp}). We will use $\gamma_\alpha(\omega)$ also when dealing with the classical and local GKLS MEs.

Lastly, we emphasize that Eq.~\eqref{eq:GKLSglob} is qualitatively different from the local ME in that the jump operators cannot be represented by a product of local terms. Moreover, since the spectrum of $H_{ab}$ contains degeneracies and some of the gaps may be repeated, the jump operators $\Lambda_\alpha(\omega)$ will generally not be of rank-$1$. Therefore, the global and local ME do not coincide even when the inter-particle interaction, controlled by $K$, goes to zero. Neither they do in the ``classical'' limit of $\tau\to 0$. The discrepancy is in fact exacerbated when $K$ or $\tau$ go to zero, which is due to the increase in the degeneracy of $H_{ab}$. Given that there is no interaction between the particles when $K\ll 1$, it is obvious that it is the description provided by the global ME that fails. Indeed, as $K$ decreases, so do also some of the gaps in the spectrum of $H_{ab}$, leading to the breakdown of the secular approximation \cite{bp, Gonzalez_2017}. In fact, in the weak coupling regime of some models \cite{Gonzalez_2017, Hofer_2017_NJP}, the local ME leads to a steady state closer to that obtained via an exact solution of the system as a whole, including the baths.

\subsection{Steady states and symmetries}
\label{sec:symmetry}

Symmetries of a system strongly affect its transport properties \cite{Magnasco_1993, Kohler_2005, Hanggi_2009, Buca_2012, Manzano_2018}. In this subsection, we will describe the symmetries of our machine's Hamiltonian and show how they relate to the transport properties of the machine.

In order to make a proper bookkeeping of the role the symmetry breaking plays in current generation, we exclude any source of asymmetry in the system-environment interaction, by not only choosing the total Hamiltonian to be symmetric under rigid rotations of the rotor as a whole, given by the operator $R = \sum_{j_a, j_b} \ket{j_a + 1, j_b + 1} \bra{j_a, j_b}$ and its integer powers, but also by additionally enforcing the particle-bath interactions to be symmetric under local rotations, given by $R_\alpha = \sum_{j_\alpha = 1}^3 \ket{j_\alpha + 1} \bra{j_\alpha}$ and its integer powers. It is easy to see that, by choosing
\bea \label{eq:As}
A_a = \frac{1}{3} X \otimes \mathbb{I}_b \quad \; \mathrm{and} \quad \; A_b = \frac{1}{3} \mathbb{I}_a \otimes X,
\eea
with $X$ as defined by Eq.~\eqref{eq:X}, we satisfy both of these symmetry properties. Being the arithmetic mean of the Gell-Mann matrices $\lambda_1$, $\lambda_4$, and $\lambda_6$ (see, e.g., \cite{Weinberg_v2}), $X$ is in fact the most natural choice for a particle-bath coupling, in that it generalizes the $x$-component of spin in $SU(2)$ (the standard $2$-dimensional Pauli $X$ operator) to $SU(3)$.

The existence of any ``strong'' symmetry of the total Hamiltonian, e.g., $[H_{tot}, R] = 0$, necessarily implies that the microscopically derived GKLS ME (see Eq.~\eqref{eq:GKLSglob}) has multiple steady states \cite{Buca_2012}. More specifically, there are at least as many linearly independent, trace-$1$ steady solutions of the ME as there are distinct eigenvalues of $R$. Importantly, this ambiguity means that the steady state of the evolution bears some memory on the initial state \cite{Buca_2012, Albert_2014}. Now, for $\phi \neq k\pi/3$, our numerical analysis reveals that the steady state depends on two free parameters. Given that the strong symmetry with respect to $R$ guarantees \cite{Buca_2012} $3$ trace-$1$ solutions to $\mathcal{L}_\mathrm{gl}[\rho]=0$, we thus conclude that each eigensubspace of $R$ contains a \textit{unique} steady state and that there are no non-zero traceless solutions to $\mathcal{L}_\mathrm{gl}[\rho]=0$. Let us now write the eigenresolution of $R$ as $R = \sum_{k=1}^3 e^{\frac{2 \pi \mathrm{i}}{3}k} R_k$, where the cubic roots of $1$ are the eigenvalues and $R_k$ are the eigneprojectors ($R_2=R_1^*$). By numerically finding an arbitrary solution for the degenerate linear system $\mathcal{L}_\mathrm{gl}[\rho]=0$---call it $\tilde{\rho}$---we determine the three mutually orthogonal steady states as $\rho^\mathrm{st}_k = \frac{R_k \tilde{\rho} R_k}{\tr(R_k \tilde{\rho})}$, which of course do not depend on $\tilde{\rho}$. An important property of the steady states is that, although they are not invariant under particle exchange, their marginals coincide for all choices of the parameters:
\bea \label{eq:equality}
\tr_a\rho^\mathrm{st}_k = \tr_b\rho^\mathrm{st}_k,
\eea
for all values of $k$.

If the rotor's initial state is $\rho_0$, then the steady state it eventually evolves into will be
\bea \label{eq:rhost}
\rho^\mathrm{st}[\rho_0] = \sum_{k=1}^3 \tr(R_k \rho_0) \rho^\mathrm{st}_k.
\eea
Indeed, keeping in mind that $\sum_k R_k=\mathbb{I}$, we can write $\rho_0 = \sum_{k} \tr(R_k \rho_0) \frac{R_k \rho_0 R_k}{\tr(R_k \rho_0)} + \sum_{k \neq k'}R_k \rho_0 R_{k'}$. Since there are no non-zero traceless solutions to $\mathcal{L}_\mathrm{gl}[\rho]=0$ and $R_k \rho_0 R_{k'}$ ($k \neq k'$) are traceless (and remain so because the evolution is trace preserving), the second sum in the decomposition will vanish as the system converges to its steady state. Whereas each of the states $\frac{R_k \rho_0 R_k}{\tr(R_k \rho_0)}$ will evolve into its corresponding $\rho_k^\mathrm{st}$, leaving us with Eq.~\eqref{eq:rhost}.

We can now use Eqs.~\eqref{eq:tuncur} and \eqref{eq:thcur} to determine the currents associated to all three $\rho^\mathrm{st}_k$'s. First of all, for all the values of the parameters, all the average currents are independent of the site (e.g., $J_{j_a\to j_a +1}[\rho^\mathrm{st}]$ is the same for all $j_a$'s). Furthermore, we find (numerically) several other general properties that also hold for all values of the parameters. First, the average current in $\rho_3^\mathrm{st}$ is zero for both particles $a$ and $b$: $J_\alpha[\rho_3^\mathrm{st}]=0$. Next, the tunneling currents of the first and second steady states are always opposite, $J^\mathrm{(tun)}_\alpha[\rho_1^\mathrm{st}] = - J_\alpha^\mathrm{(tun)}[\rho_2^\mathrm{st}]$, while their thermal currents always coincide: $J_\alpha^\mathrm{(th)} [\rho_1^\mathrm{st}] = J_\alpha^\mathrm{(th)} [\rho_2^\mathrm{st}]$. Lastly, due to Eq.~\eqref{eq:equality}, we also have $J_a^\mathrm{(tun)}[\rho^\mathrm{st}_k] = J_b^\mathrm{(tun)}[\rho^\mathrm{st}_k]$, for any $k$. In view of all this, and using the numerical values of $R_{1}$ and $R_2$, for an arbitrary initial state $\rho_0$, we find
\bea \label{eq:Barbados}
J[\rho^\mathrm{st}[\rho_0]] = \frac{2 \, \mathrm{Im}\,\theta_0}{\sqrt{3}} J^\mathrm{(tun)}[\rho^\mathrm{st}_1] + \frac{2-2 \, \mathrm{Re}\,\theta_0}{3} J^\mathrm{(th)}[\rho^\mathrm{st}_1], ~ 
\eea
where $\theta_0 = \sum_{j_a,j_b=1}^3 \bra{j_a,j_b} \rho_0\ket {j_a + 1, j_b + 1}$ is the sum of some of the non-diagonal elements of the initial density matrix.

\begin{figure}[t!]
\centering
\includegraphics[width=0.48\textwidth]{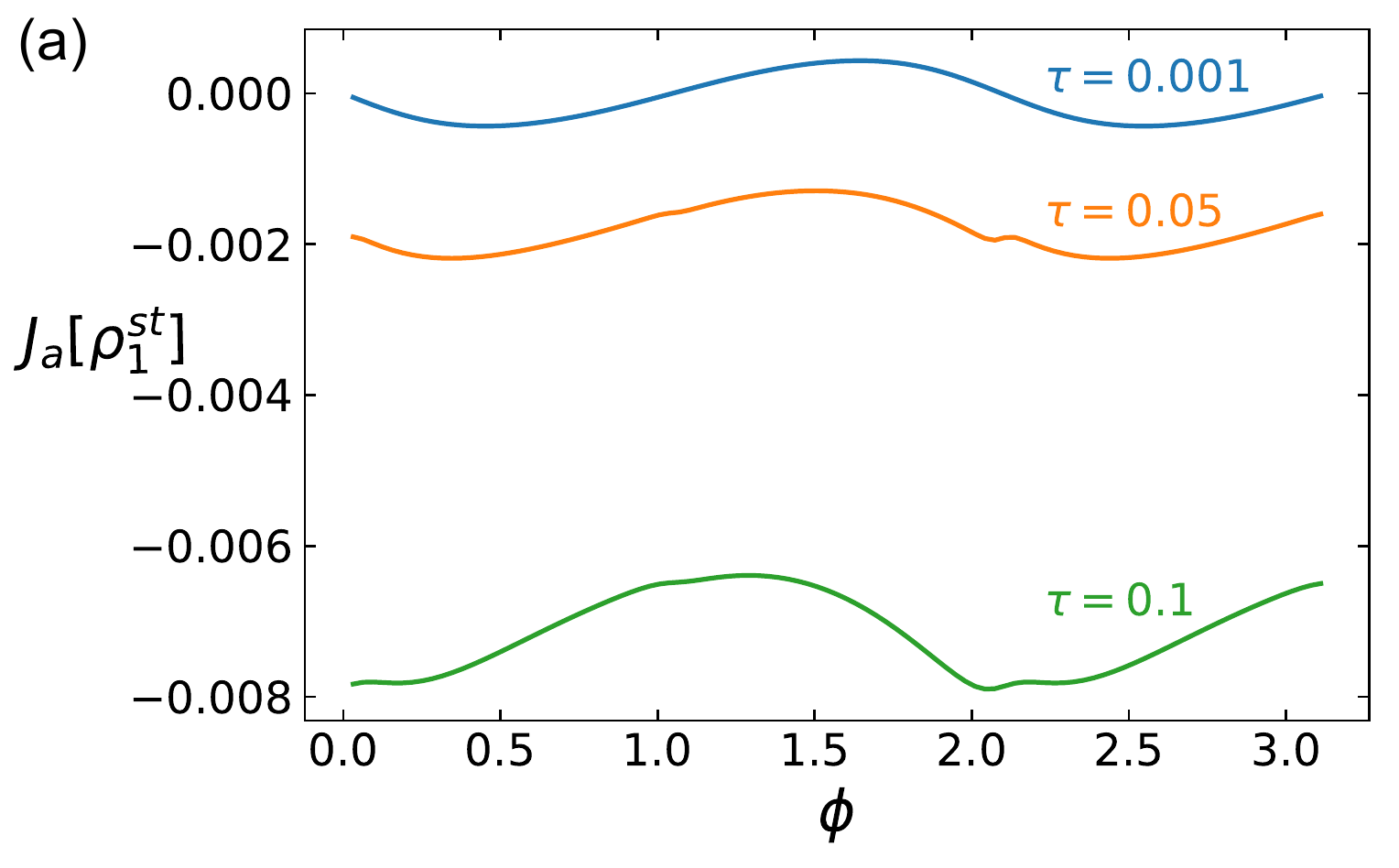}
\includegraphics[width=0.48\textwidth]{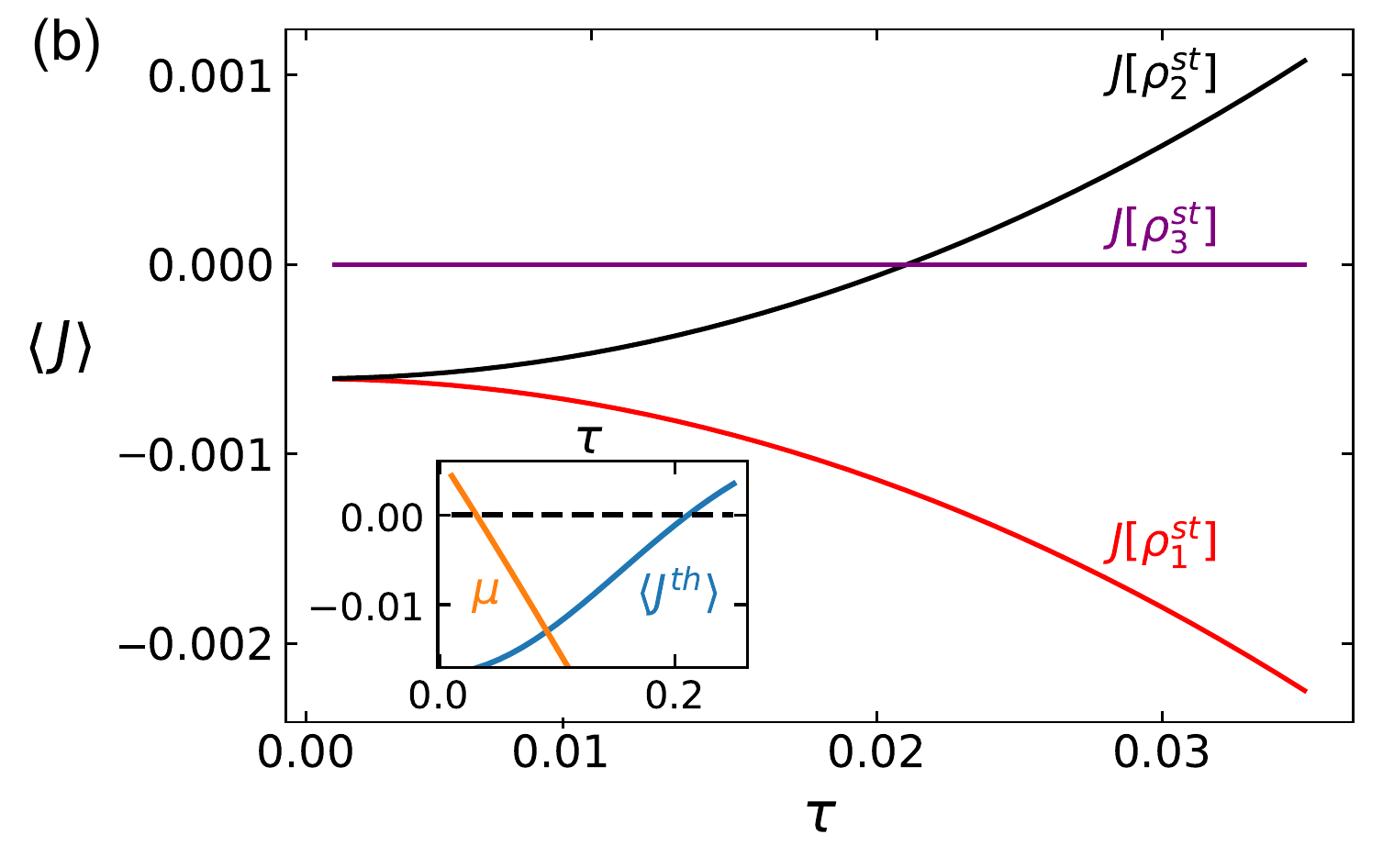}
    \caption{Current. \textbf{(a)} The current of particle $a$ versus the phase $\phi$ in the first steady state $\rho^\mathrm{st}_1$, for different values of the tunneling coefficient $\tau$. The blue curve corresponds to near-zero tunneling ($\tau=0.001$), and the orange and green curves correspond to, respectively, $\tau=0.05$ and $\tau=0.1$. \textbf{(b)} The total average current versus the tunneling rate $\tau$, with the phase fixed to $\phi=\pi/6$. The total current starts negative, and, depending on the initial state, can turn positive as $\tau$ increases. The \textbf{(inset)} shows the average thermal current of particle $a$ in the first steady state, $\langle J^{th}_a[\rho_1^\mathrm{st}] \rangle$ ($ \times 10^2$), and the time derivative of the Margenau-Hill distribution, $\mu_a$ (see the text), as a function of $\tau$. It is clearly seen that the weak value, which has the same sign as $\mu_a$, turns negative before the current changes its sign. For both plots, the rest of the parameters are: $T_a = 0.2$, $T_b = 1$, $\gamma_a = \gamma_b = 0.2$, $K = 2$.}
\label{fig:fig2}
\end{figure}

Importantly, the average steady-state current is a $2\pi/3$-periodic function of $\phi$. This fact can be appreciated already in Fig.~\ref{fig:fig2}a, and is rigorously proven in \ref{app:period}.

Another important feature of the global GKLS ME is that, whenever $\phi = k\pi/3$ ($k$ is an arbitrary integer), $J^\mathrm{(th)} = 0$ for any initial state. This nullification of the thermal current also takes place for both the classical and phenomenological quantum MEs (see \ref{app:symm}), and is related to the presence of an exchange symmetry, which is broken whenever $\phi \neq k \pi / 3$. More specifically, if $\phi = k\pi/3$, then $U_{j,i} = U_{i, j + k}$; i.e., the potential is symmetric under the combination of particle exchange and unilateral rotation. The same symmetry applies also to the quantum Hamiltonian of the rotor, $H_{ab}$. Indeed, the unitary operator corresponding to the generalized exchange is given by $\Xi_k = \mathrm{SWAP} \cdot \left( \mathbb{I} \otimes R_b^k \right)$, where $\mathrm{SWAP}$ is the swap operator (for $\forall$ $\ket{\psi}$ and $\ket{\xi}$, $\mathrm{SWAP} \ket{\psi \xi}=\ket{\xi \psi}$), and it is easy to see that $[H_{ab}, \Xi_k]=0$. In other words, in order to generate thermal current, one needs to break the generalized exchange symmetry of the rotor Hamiltonian. It is this asymmetry that provides the ratchet effect propelling the rotor both in the classical and quantum regimes. In Fig.~\ref{fig:fig2}a, we plot the total (thermal plus tunneling) current of particle $a$ against the phase $\phi$, for different values of $\tau$. We see that, in contrast to the thermal current, the tunneling current is not zero when $\phi=k\pi/3$, which is the case only for the global ME. Notice also that the average current for nearly zero tunneling rate is noticeably smaller than the average currents for larger tunneling rates.

When $\phi=k\pi/3$, the Hamiltonian of the rotor is symmetric under $\Xi_k$, whereas the coupling operators, $A_a$ and $A_b$ (and therefore also $H_\mathrm{tot}$), are not. Nevertheless, $\phi=k\pi/3$ is marked by an increased degeneracy in the spectrum of $H_\mathrm{tot}$, and, as a consequence, of $\mathcal{L}_\mathrm{gl}$. This is expressed in the fact that, although not invariant under $\Xi_k$, the steady state now depends on $5$ free parameters, meaning that the steady-state space of $\mathcal{L}_\mathrm{gl}$ is $6$-dimensional. This extra symmetry is responsible for the nullification of the thermal current. In \ref{app:symm}, we provide an extended discussion of these issues for the phenomenological quantum and classical MEs.

An interesting purely quantum effect can be read off from Fig.~\ref{fig:fig2}b (in which we plot the total average current for all three basis steady states, as a function of $\tau$): the current changes its direction as the tunneling rate increases. Keeping Eqs.~\eqref{eq:rhost} and \eqref{eq:Barbados} in mind, we see that, unless the initial state is completely in the subspace of $R_3$, $\langle J \rangle <0$ for $\tau\ll 1$, and if $\mathrm{Im}\,\theta_0<0$, $\langle J \rangle$ will become positive for large enough $\tau$. Such a reversal of the current is somewhat counterintuitive in that the tunneling is symmetric with respect to both global and local rotation, and it therefore does not favour a particular direction for particle flow. This is a purely quantum phenomenon, occurring only when $T_a$ is sufficiently low. In fact, it turns out that the inversion of the thermal current is always preceded, implying (yet by no means proving) a causal relation, by the the onset of measurement-contextuality in the system, marked by the weak value in Eq.~\eqref{eq:weakval} turning negative (see the related discussion in Sec.~\ref{sec:current} and \ref{app:context}). We illustrate this in the inset of Fig.~\ref{fig:fig2}b, where we show both $\langle J^{th}_a[\rho_{1}^\mathrm{st}] \rangle$ and $\mu_{j_a, j_a+1} = \lim\limits_{\epsilon \to + 0} \frac{1}{\epsilon}\mathrm{Re} \, \tr [x_{j_a} \, x_{j_a+1} (\epsilon) \rho_{1}^\mathrm{st}]$ against $\tau$. We chose to plot $\mu_{j, j'}$---the time-derivative of the Margenau-Hill distribution (see Sec.~\ref{sec:current})---instead of the weak value since not only $\mu_{j, j'}$ has the same sign as the weak value (cf. Eq.~\eqref{eq:weakval}), but it also directly appears in the expression for the average current: $\langle J_{j \to j'} \rangle = \mu_{j, j'} - \mu_{j', j}$ (see Eq.~\eqref{eq:avweakcur}). Let us note that current inversion, albeit through the fundamentally different mechanism of external periodic driving, was also reported in Brownian ratchet systems \cite{Linke_1999, Reimann_2002}.

Lastly, we note that, whenever $T_a = T_b$, $J^\mathrm{(th)}_\alpha = 0$ for any initial state. The tunneling currents, on the other hand, do not necessarily turn to zero when $T_a = T_b$. This however does not contradict the second law of thermodynamics, and is simply due to the fact that, although the thermal state---for which the tunneling currents are indeed zero---is a steady-state solution for the global ME, the basis steady states, $\rho_k^\mathrm{st}$, are not thermal.

\subsection{Heat Flux} \label{sec:heat}

In the absence of external driving, heat always flows from hot to cold. Due to clear separation between the two baths in the total Hamiltonian $H_\mathrm{tot}$, the contributions of different baths in the resulting GKLS ME are simply summed (this is true also for the phenomenological ME). Therefore, casting the generator of the evolution in the form $\mathcal{L}_\mathrm{gl} [\rho] = - \mathrm{i} [H_{ab}, \rho] + \mathcal{D}_a [\rho] + \mathcal{D}_b [\rho]$, where $\mathcal{D}_\alpha[\rho] = \sum_\omega \gamma_\alpha(\omega) \mathcal{S} [\Lambda_\alpha (\omega)] [\rho]$, the standard definition of the heat flow yields
\bea \label{eq:Qdot}
\dot{Q}_\alpha = \tr (H_{ab} \mathcal{D}_\alpha [\rho]).
\eea
Since the system does not exchange energy with external media, in the steady state, the energy conservation requires $\dot{Q}_{a} + \dot{Q}_{b} = 0$. Moreover, at thermal equilibrium, i.e., when $T_a = T_b$, the heat flow must be zero. The first condition is trivially satisfied by any GKLS equation, whereas, in general, the second condition is guaranteed only when one uses the global ME (although, in our model, this is also respected by the phenomenological ME).

\begin{figure}[t!]
\centering
\includegraphics[width=0.65\textwidth]{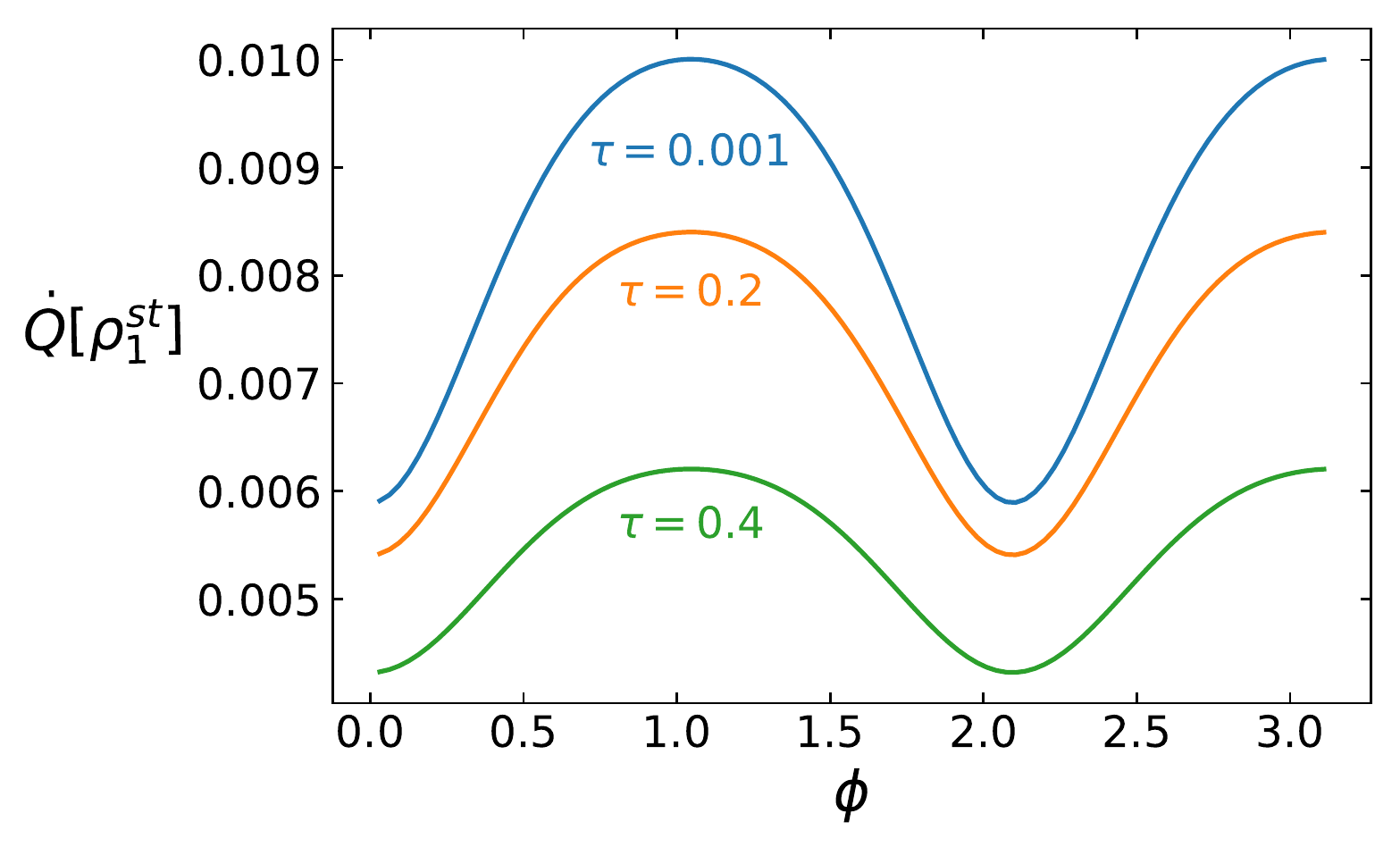}
    \caption{Heat. The steady-state heat flux from hot to cold reservoir, versus $\phi$, for different values of the tunneling coefficient $\tau$. The blue curve corresponds to near-zero tunneling ($\tau=0.001$), and the orange and green curves correspond to, respectively, $\tau=0.2$ and $\tau=0.4$. The rest of the parameters are the same as those for Fig.~\ref{fig:fig2}.}
\label{fig:fig3}
\end{figure}

We have numerically established the following properties of the heat flow: $\dot{Q}[\rho^\mathrm{st}_1] = \dot{Q}[\rho^\mathrm{st}_2] > \dot{Q}[\rho^\mathrm{st}_3]>0$ and $\dot{Q}[\rho^\mathrm{st}_1] - \dot{Q}[\rho^\mathrm{st}_3]\propto \tau$ (for $\tau\ll 1$). This means that, when the tunneling is not zero, the heat flow can be controlled by changing the weights of the eigensubspaces of $R$ in the initial state of the rotor; such symmetry-controlled manipulation is well-known in the literature \cite{Manzano_2018}. In Fig.~\ref{fig:fig3}, we plot $\dot{Q}[\rho^\mathrm{st}_1]$ ($\dot{Q}[\rho^\mathrm{st}_3]$ has the same behavior) against $\phi$, for various values of $\tau$. First, we notice that the heat flow decreases with increasing the tunneling rate, which is related to the fact that the higher the $\tau$, the higher the relative weight of the non-interacting component of $H_{ab}$. We further notice that the zeroes of $J^{(th)}$, i.e., $\phi=k\pi/3$, correspond to global extrema of $\dot{Q}$. Lastly, analogously to the current, the heat flow is also a $2\pi/3$-periodic function of $\phi$ (cf. Fig.~\ref{fig:fig3} and see \ref{app:period} for a proof).

We conclude this subsection by emphasizing that caution must be maintained when using the global ME for computing the heat flow for small $K$. As discussed above, the secular approximation, indispensable for the applicability of the global ME, is compromised when $K$ approaches $0$. Given the results in Refs.~
\cite{Gonzalez_2017, Hofer_2017_NJP} for the Caldeira-Leggett model, it is reasonable to expect that the local ME would provide a more reliable description in that regime. This issue can be decisively settled only upon solving the global, system-plus-baths dynamics in the limit of infinitely large baths, doing which, however, appears to be unfeasible \cite{bp}.

\subsection{Steady-State Entanglement}

Entanglement is a key resource in basically all aspects of quantum technology \cite{mikeike, 4H_2009}. Although thermal noise can sometimes be beneficial for creating and maintaining entanglement \cite{Braun_2002, Sinaysky_2008, Bellomo_2013}, at high temperatures, the effect of environment will typically be detrimental. Indeed, the maximally mixed (or, equivalently, infinite-temperature) state in any finite-dimensional Hilbert space is not entangled, and there exists a finite-volume convex set around it, in which all states are not entangled \cite{4H_2009}. In particular, this means that, for any Hamiltonian, $H$, there exists a finite temperature $T[H]$, above which all thermal states are non-entangled.

On the other hand, thermal disequilibrium is known to be beneficial for entanglement generation \cite{Quiroga_2007, Sinaysky_2008, Kheirandish_2010, Scala_2011, Wu_2011, Bellomo_2013, Brask_2015_NJP, Tavakoli_2018}. Therefore, a key question in this regard is whether a thermal disequilibrium created by temperatures higher than $T[H]$ can generate entangled non-equilibrium states. Of especial interest would be the steady states, since these can be prepared and maintained reliably and robustly. Here, we will show that a thermal initial state can be transformed into an entangled steady state only if it is already entangled. Thus, for thermal initial states, the above question has a negative answer. However, it turns out that, for initially uncorrelated states, local coherence can be traded for entanglement in the steady state, for \textit{any} temperatures of the baths. In other words, powered by global rotation symmetry, our machine can convert a local quantum resource (coherence) into a global quantum resource (entanglement), even when the surroundings have such high temperatures that neither resource would survive thermalization.

Our rotor is a $3\times 3$-dimensional system, and in $9$ (or higher) dimensions it is generally very hard to tell whether a state is entangled or not \cite{4H_2009}. In this paper, we will use a very strong \textit{necessary} condition, called positive partial transpose (PPT) criterion \cite{4H_2009}. The partial transpose of $\rho$ with respect to partition $b$, $\rho^{\Gamma_b}$, is defined as the matrix with entries $\bra{j_a, j_b}\rho^{\Gamma_b}\ket{j_a',j_b'} = \bra{j_a, j_b'} \rho \ket{j_a', j_b}$. Now, if $\rho$ is not entangled, then $\rho^{\Gamma_b}$ will be a non-negative matrix. Hence, $\rho$ will necessarily be entangled if $\rho^{\Gamma_b}$ has a negative eigenvalue. However, in dimensions higher than $6$, e.g., for our rotor, this condition is not sufficient: $\rho$ can be entangled even when $\rho^{\Gamma_b}\geq 0$ (it is then said to be ``bound entangled'' \cite{4H_2009}). The entanglement detected by the PPT criterion can be measured by the \textit{negativity},  $\mathcal{N}[\rho]$, defined as the sum of the absolute values of all negative eigenvalues of $\rho^{\Gamma_b}$ \cite{Zyczkowski_1998}: $\mathcal{N}(\rho) = \frac{1}{2} \left(|| \rho^{\Gamma_b} ||_1 - 1 \right)$, where $|| O || = \tr \sqrt{O^\dagger O}$ is the trace norm \cite{mikeike}.

\begin{figure}[t!]
\centering
\includegraphics[width=0.65\textwidth]{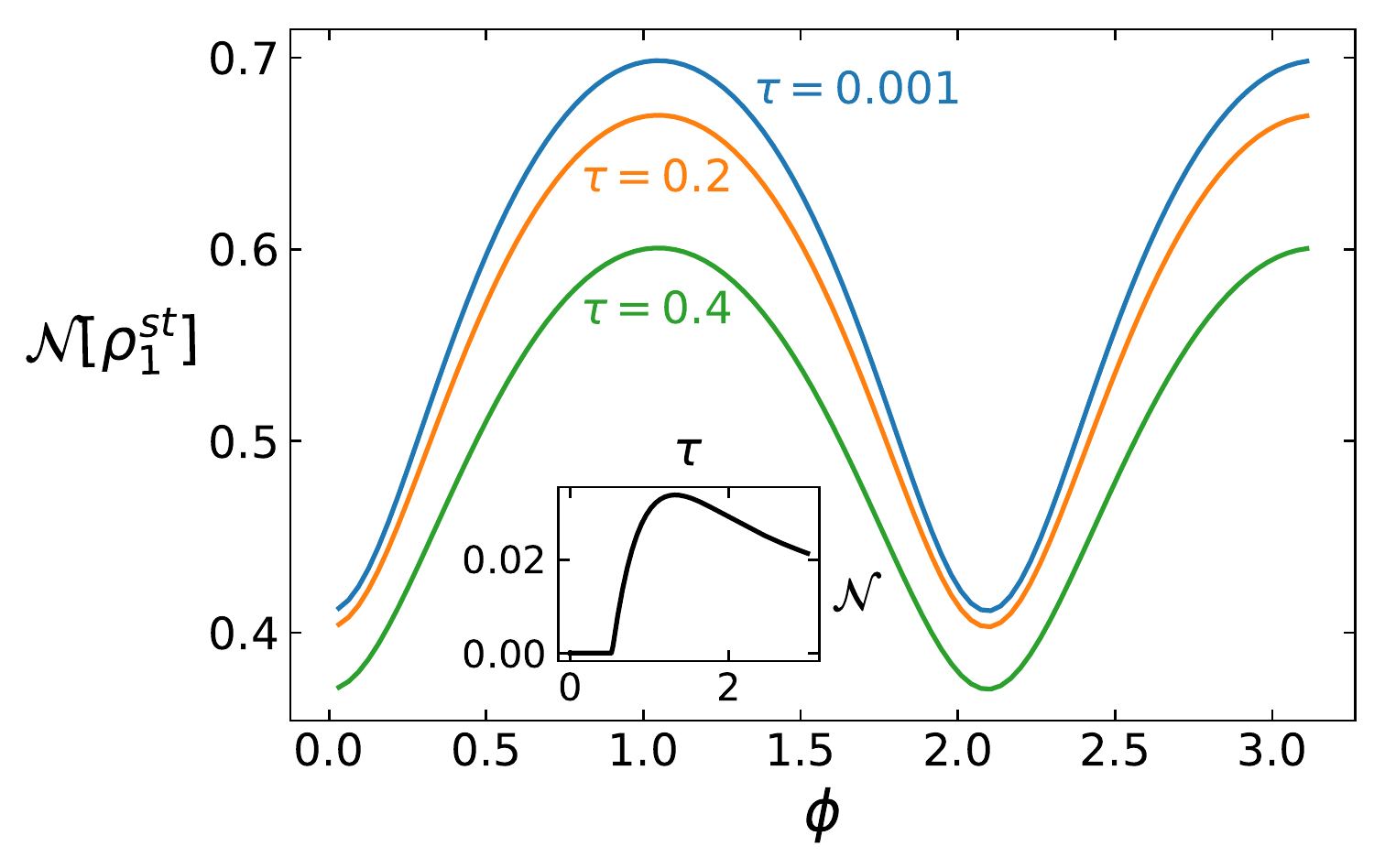}
    \caption{Entanglement. The steady-state entanglement between particles $a$ and $b$, as measured by the negativity, versus the phase $\phi$. Only the negativity for $\rho_1^\mathrm{st}$ (which is also equal to that of $\rho_0^\mathrm{st}$) is plotted since $\mathcal{N}(\rho_2^\mathrm{st}) < \mathcal{N}(\rho_1^\mathrm{st})$ has the same behavior. The \textbf{(inset)} shows the steady-state negativity when the initial state is chosen to be the thermal state at the temperature of the hot bath, as a function of $\tau$, with $\phi=\pi/3$. The rest of the parameters are the same as those in Fig.~\ref{fig:fig2}.}
\label{fig:fig4}
\end{figure}

Similarly to the heat flux, we find numerically that (i) $\mathcal{N}(\rho^\mathrm{st}_1) = \mathcal{N}(\rho^\mathrm{st}_2) > \mathcal{N}(\rho^\mathrm{st}_3)>0$ and (ii) the global extrema of the negativity correspond to the zeroes of the thermal current ($\phi = k\pi/3$). In Fig.~\ref{fig:fig4}, we plot $\mathcal{N}(\rho^\mathrm{st}_1)$ against $\phi$, for three different values of tunneling: $\tau=10^{-3}$, $\tau=0.2$, and $\tau=0.4$. It shows, in particular, that the negativity is a monotonically decreasing function of $\tau$. This seemingly counterintuitive phenomenon can be easily understood by first observing that the presence of entanglement even when $\tau \to 0$ is caused by the fact that, in order to obtain the basis steady state, we project on the highly entangled eigensubspaces of the global rotation matrix $R$; and the decrease of entanglement with the increase of $\tau$ is caused by the increased weight of the non-interacting part of the Hamiltonian, $H_{ab} - H_{int}$ (see Eq.~\eqref{eq:Haml}). The standard intuition is recovered when one considers realistic initial states. We show this in the inset of Fig.~\ref{fig:fig4}, where we plot the steady-state negativity versus $\tau$. There, we fix $K = 2$ and $\phi = \pi/3$, and, for any $\tau$, choose the initial state to be the thermal state at the temperature of the hot bath: $\rho_0 \propto e^{-\beta_b H_{ab}}$. As expected, we see that there is no entanglement for weak tunneling. Then, having reached a maximum at an intermediate value of $\tau$, the entanglement decays as $H_{ab}$ becomes more and more dominated by the non-interacting tunneling term.

Note also that, as is visible in Fig.~\ref{fig:fig4}, similarly to the current and heat flux, the negativity is a $2\pi/3$-periodic function of $\phi$. See \ref{app:period} for the proof.

The dependence of the steady-state entanglement of global GKLS MEs on the temperature difference of the baths has been extensively studied in qubit systems \cite{Quiroga_2007, Sinaysky_2008, Kheirandish_2010, Bellomo_2013, Scala_2011}. In highly asymmetric problems, situations when entanglement can be generated only when there is temperature difference have been reported \cite{Bellomo_2013}. In our model, we find that, for any values of all the other parameters fixed, if we set the cold temperature $T_a$ constant, then the negativity of all three basis steady states is a monotonically decreasing function of $T_b$. Furthermore, the negativity at $T_b=T_a$ is a monotonically decreasing function of $T_a$. Interestingly, however, we find that, when both keeping $T_a$ constant and varying $T_b$ or fixing $T_b = T_a$ and varying $T_a$, the entanglement does not experience sudden death: the negativity decays gradually, reaching zero only asymptotically. Therefore, by choosing the initial state appropriately, we can guarantee the presence of entanglement in the steady state for arbitrarily large values $T_a$ and $T_b$. More importantly, we can achieve this with uncorrelated initial states, albeit paying for that with local coherence. Indeed, as we prove in \ref{app:titfortat}, for any $T_a$ and $T_b$, there exist two single-particle states, $\sigma_a$ and $\sigma_b$, such that the machine evolves $\sigma_a \otimes \sigma_b$ into an entangled steady state. As $\min \{ T_a, T_b \}$ increases, $\sigma_a$ and $\sigma_b$ become monotonically less mixed and more coherent. In the limit of infinite temperatures, $\sigma_a$ and $\sigma_b$ necessarily become pure and maximally coherent (with respect to the measure of coherence defined as $\mathcal{C}[\sigma] = \sum_{j \neq j'}|\sigma_{jj'}|$ \cite{Baumgratz_2014}). The family $\sigma_\alpha(\delta) = (1 - \delta) \ket{\sigma_\alpha}\bra{\sigma_\alpha} + \frac{\delta}{3} \mathbb{I}$, with $\ket{\sigma_a} = \frac{1}{\sqrt{3}} \left( \ket{1} + e^{-2\pi\mathrm{i}/3} \ket{2} + e^{2\pi\mathrm{i}/3} \ket{3} \right)$ and $\ket{\sigma_b} = \frac{1}{\sqrt{3}}\left( \ket{1} + \ket{2} + \ket{3} \right)$, represents an example of such states: for any $T_a$ and $T_b$, there exists $\delta >0$ ($\delta \to 0$ as $\min \{ T_a, T_b \} \to \infty$) such that $\rho^\mathrm{st}[\sigma_a(\delta) \otimes \sigma_b (\delta)]$ is entangled.

\subsection{Work Content}

\begin{figure}[t!]
\centering
\includegraphics[width=0.65\textwidth]{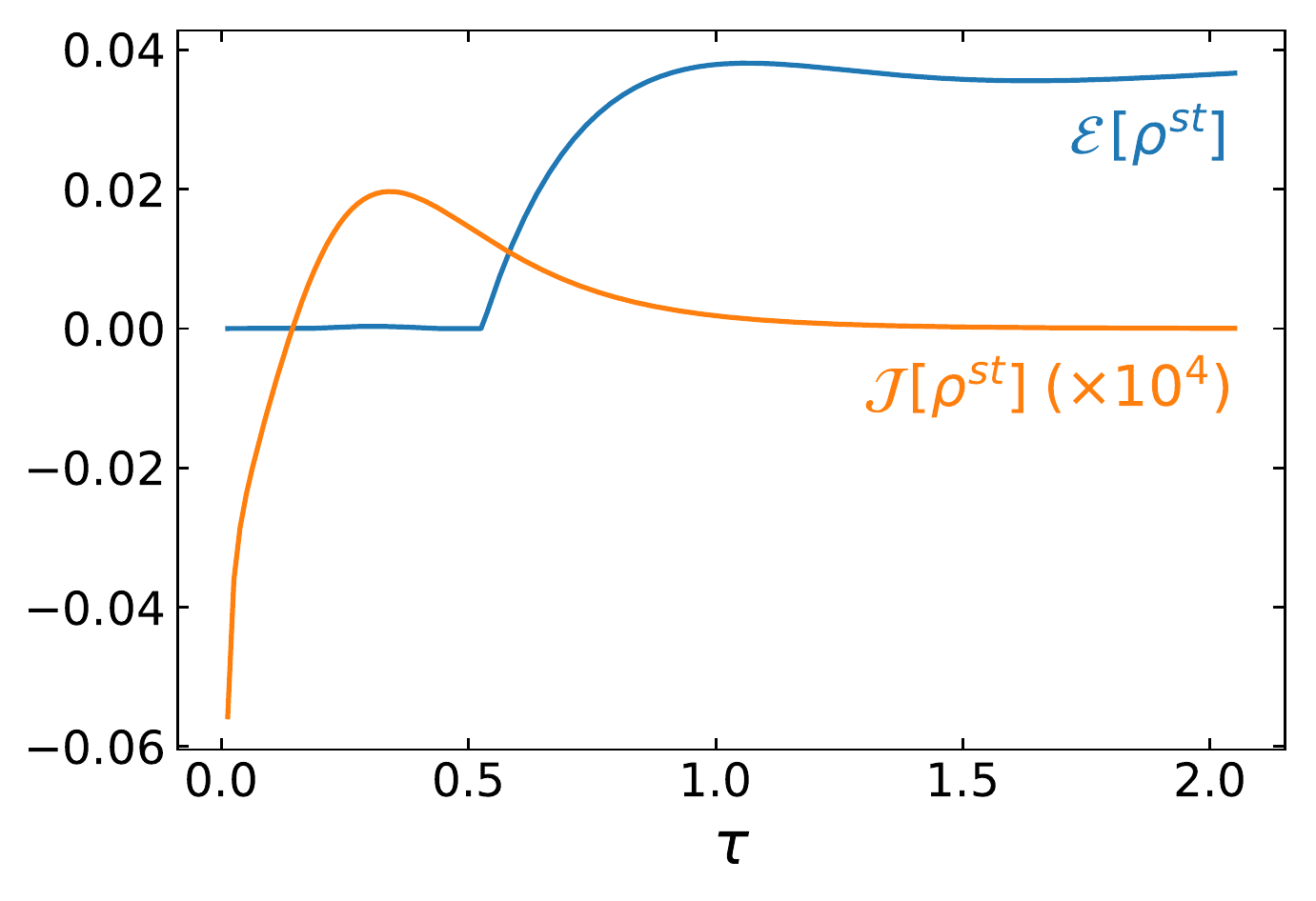}
    \caption{Ergotropy. The ergotropy of the steady state when the initial state is a thermal state at the temperature of the cold bath, versus the tunneling coefficient $\tau$. The plot also shows the total current in order to show that the amount of work stored in the state is not related to the current flowing in it. As an aside, the plot for the current once again illustrates the phenomenon of current inversion. The other parameters are the same as those in Fig.~\ref{fig:fig2}.}
\label{fig:fig5}
\end{figure}

The steady states of our machine can be useful not only in quantum information, but also in thermodynamics. In particular, the non-equilibrium steady states of the rotor store work that can be extracted from it in a cyclic Hamiltonian process. In other words, the average energy of the rotor, with respect to its Hamiltonian $H_{ab}$, can be decreased by unitarily evolving its state $\rho^\mathrm{st}$. The unitary operation that extracts maximal work is the one that diagonalizes $\rho^\mathrm{st}$ in the basis of $H_{ab}$ in such a way that, if $H_{ab}$'s eigenbasis is chosen so that $E_1\leq E_2 \leq \cdots$, then the eigenvalues of $\rho^\mathrm{st}$, $\{ r_k \}$, have the reverse ordering: $r_1 \geq r_2 \geq \cdots $ \cite{Allahverdyan_2004}. The average work thus extracted, $\mathcal{E}$, is called ergotropy \cite{Allahverdyan_2004} and is equal to
\bea
\mathcal{E}[\rho^\mathrm{st}] = \tr(H_{ab} \rho^\mathrm{st}) - \sum_m E_m r_m.
\eea
The states with zero ergotropy are called passive \cite{Pusz_1978, Allahverdyan_2004}, with the most prominent example being the thermal states \cite{Pusz_1978}.

When the rotor's initial state is thermal at temperature $T_{in}$, the ergotropy of the steady state will be a decreasing function of $T_{in}$, turning to zero somewhere in between $T_a$ and $T_b$. Now, given the infinite size of the baths, thermal states at the temperatures of the baths can be considered to be for free---one just puts the system in contact with the bath and waits for it to thermalize. Therefore, if $T_{in}=T_a$, the machine essentially takes in a free (and useless from the perspective of work extraction) state and evolves (and maintains for an arbitrarily long time) it into a state ``charged'' by work. In Fig.~\ref{fig:fig5}, we plot $\mathcal{E} \left[\rho^\mathrm{st} \left[e^{-H_{ab}/T_a}/Z\right]\right]$, where $Z = \tr e^{-H_{ab}/T_a}$ is the partition function, as a function of $\tau$. We see that, in addition to a temperature gradient, a sufficiently high tunneling rate is necessary for the machine to work as a charger. The plot also shows the total steady-state current, in order to indicate the surprising fact that the work content of the steady state is not related to (let alone being caused by) the current it harbours.

Note that non-zero temperature gradient, necessary to charge an initially thermal rotor, is not needed in general: for most values of the parameters, $\rho^\mathrm{st}_i$ are all non-passive states. Therefore, by suitably choosing the initial state, one can obtain a non-passive steady state even when $T_a = T_b$. In such cases, the ergotropy of the initial state is always larger or equal to that of the resulting steady state. Importantly, there will be no heat current in the steady state, meaning that the maintenance of the non-passive steady state will be energetically free. Albeit via a different mechanism, a very similar effect---a system interacting with a single thermal bath reaching a non-passive steady state with zero steady-state heat flow---has been recently reported in Ref.~\cite{Barra_2019}.

\section{Conclusions}

In this work, we have derived a general formula (Eq.~\eqref{eq:current}) for the current of a quantum particle on an arbitrary graph, undergoing an arbitrary open-system dynamics. Our formula recovers all known definitions of probability/particle current \cite{Messiah_v1, Gebauer_2004, Kohler_2005, Bodor_2006, Zhang_2008, Avron_2012}, in the corresponding regimes of their validity. In the classical limit, it reduces to the standard Eq.~\eqref{eq:clacur}. Usable in all situations, our formula describes the current even when the position operator does not commute with the system-environment interaction, which, until now, was an uncharted territory. We further deepen the formalism by providing an interpretation for our formula in terms of quantum weak measurements. We show that the average current is a result of two position measurements---one weak and one strong. This sheds new light also on well-known and widely-used formulas of particle current, such as, e.g., Eq.~\eqref{eq:tuncur}.

As a next step, we used the developed formalism to study current generation on a minimal, two-particle toy model, autonomously operating between two thermal baths. There, by carefully accounting for all relevant spatial symmetries of the full system-environment Hamiltonian, we established that the breaking of particle exchange symmetry governs current generation both classically and deep in the quantum regime. Counterintuitively, the model exhibits reversal of the direction of the steady-state current as the rotationally invariant tunneling rate is varied. We argued that this is a purely quantum effect by showing that, in our model, it appears only when the system exhibits quantum contextuality.

We also showed that our device can be useful in such practical tasks as converting uncorrelated states into entangled states in the presence of arbitrarily hot dissipative environment and evolving initially thermal states, which are useless for work extraction, into non-passive steady states, from which non-zero work can be extracted. Both tasks are enabled by purely quantum resources. The entanglement generation is made possible by feeding the machine with coherent local states. In other words, the evolution to the steady state is a realization of a completely positive trace preserving map, converting local coherence into entanglement which depends on the amount of coherence at the input, akin to the analysis in Ref.~\cite{Regula_2018}. In the second case, ``charging'' of the initially thermal rotor is possible only if the tunneling rate is sufficiently high.

The presented theory unlocks the possibility of describing particle current in most general physical scenarios, and, in particular, to devise practical quantum thermal motors without the limitation of having the rotor isolated from the environment. It may also open a new perspective on the study of coherent energy \cite{Iles-Smith_2016} and information transport \cite{Patel_2017} in open quantum systems undergoing general non-Markovian dynamics.

The facts that particle current can witness the onset of quantum contextuality and noise-resistant entanglement can be produced in our simple model, add constructively to the ongoing discussion on the genuinely quantum effects in the thermodynamic context \cite{Goold_2016, Lostaglio_2018} and the practical usefulness of nanoscale thermal machines \cite{Goold_2016, Bellomo_2013, Tavakoli_2018}.

\section*{Acknowledgements}

We thank Matteo Lostaglio for insightful discussions about contextuality and for pointing us to Refs.~\cite{Pusey_2014, Lostaglio_2018}. We also thank Felipe Barra for useful comments on an early draft. This work was supported by the Danish Council for Independent Research and by the Villum Foundation.

\appendix

\section{Open-system Heisenberg picture}
\label{app:Heis}

When the dynamics is unitary, the Heisenberg picture is used to shift the time dependence from the state to the operators (see any textbook \cite{Messiah_v1} on quantum mechanics): the pair $(\rho(t), O )$, where $O$ is an observable and $\rho(t) = U(t) \rho U(t)^\dagger$ is the time-dependent state, is taken to $( \rho, O(t) )$, where $O(t) = U(t)^\dagger O U(t)$. The two pictures are equivalent in that $\tr[\rho(t) O] = \tr[\rho O(t)]$ for any $t$ and any $O$, and the ultimate convenience of the Heisenberg picture stems from the simple fact that it preserves the algebra of observables: $(O_1 O_2)(t) = O_1(t) O_2 (t)$.

The situation is more complicated for non-unitary dynamics. For simplicity, let us focus on the case when the evolution, $\rho(t) = \mathcal{E}_t [\rho]$, is given by a completely positive, trace-preserving (CPTP) map \cite{mikeike}: $\mathcal{E}_t[\rho] = \sum_\mu K_\mu(t) \rho K_\mu^\dagger (t) $, where $K_\mu$---the so-called Kraus operators---satisfy $\sum_\mu K_\mu^\dagger (t) K_\mu (t) = \mathbb{I}$. Here, as above, the Heisenberg picture is defined through the requirement of $\tr(\rho(t) O) = \tr (\rho O(t))$, for any $\rho$, which yields $O(t) = \sum_\mu K_\mu^\dagger (t) O K_\mu (t)$. However, unlike in the unitary case (i.e., when there is more than one non-zero Kraus operator), the algebra of observables is not preserved: $(O_1 O_2)(t) \neq O_1(t) O_2(t)$. When the evolution is not given by a CPTP map, the transition to the Heisenberg picture, although still determined through requiring the averages to be the same, will have to be done on an ad hoc basis, and one will generally be even further from preserving the observable algebra as one was in the case of CPTP evolution. Nevertheless, one general statement can be made: the identity operator remains unchanged upon transitioning to the Heisenberg picture. Indeed, any quantum dynamics, completely positive or not, preserves Hermiticity, positivity, and trace. The latter is thus equivalent to the preservation of the identity operator.

Coincidentally, the requirement that the transition to the Heisenberg picture in Eq.~\eqref{eq:current} has to be done at the same moment of time as the current is calculated, exempts the operator products in Eq.~\eqref{eq:current} from the problem of the Heisenberg picture not preserving the observable algebra. Indeed, all the operators in Eq.~\eqref{eq:current} are calculated at the zeroth second of the Heisenberg picture, meaning that $x_j$ and $x_{j'}$ coincide with their Shr\"{o}dinger-picture values. When it comes to the derivatives of $x_j$ and $x_{j'}$, the situation is tractable when the dynamics is given by a CPTP map which is divisible \cite{Wolf_2008} at the moment of time $t$, which is when we are calculating the current. Indeed, the divisibility means that, for any $\epsilon>0$, the decomposition $\mathcal{E}_{t+\epsilon} = \mathcal{E}_t \circ \mathcal{C}_\epsilon$, where $\mathcal{C}_\epsilon$ is a CPTP map, holds. Let $\kappa_\mu (\epsilon)$ be the Kraus operators of $\mathcal{C}_\epsilon$. Then, $x_j(\epsilon) = \sum_\mu \kappa_\mu^\dagger x_j \kappa_\mu$, whence we can calculate $\left. \frac{d x_j(\epsilon)}{d \epsilon} \right|_{\epsilon = 0}$, which is the derivative required in Eq.~\eqref{eq:current}. When the dynamical map is not divisible, e.g., when the evolution is non-Markovian, the time-derivative of $x_j$ may depend on the past of the dynamics, which can make the calculation of the current intractable. A tractable class of non-Markovian processes are the time-local master equations, e.g., non-secular Redfield equations \cite{bp}, where $\left. \frac{d x_j(\epsilon)}{d \epsilon} \right|_{\epsilon = 0}$ can be calculated at any moment of time without the need to have access to quantities at previous times.

Quantum master equations of the GKLS form (see Eq.~\eqref{eq:GKLS}) constitute an important class of divisible maps \cite{bp, Wolf_2008}. When the dynamics is described by a GKLS master equation, one can readily bypass the Kraus representation and work directly with the Heisenberg equation \cite{bp}, Eq.~\eqref{eq:tuncur}, irrespective of the moment of time at which one transitions from the Schr\"{o}dinger picture to that of Heisenberg. Therefore, the particle current operator, as defined by Eq.~\eqref{eq:current}, is time independent, and is given by Eqs.~\eqref{eq:tuncur} and \eqref{eq:thcur}. Note that the GKLS equation holds only on the coarse-grained time scale where the fast environmental degrees of freedom are averaged out \cite{bp}. For example, this means that the limit $0< \epsilon \ll 1$ in Eq.~\eqref{eq:avweakcur} is to be understood as $\epsilon$ being much smaller than the time-scale of the system evolution, but much larger than the relaxation time-scale of the environment.

\section{Continuous limit} \label{app:conti}

In order to understand how our formalism extends to the continuous space, let us (a) take a continuous space and a Hamiltonian in it, (b) discretize the space and the Hamiltonian in such a way that all relevant operators return to their continuous values as the discretization step is taken to zero, (c) calculate the current $J_{j \to j'}$ using Eq.~\eqref{eq:current}, and (d) revert to the continuous limit by taking the discretization unit to zero.

We illustrate this on the generic example of a thermally isolated one-dimensional particle, with mass $m$, subject to some potential $V$:
\bea
H_c = \frac{\hat{p}^2}{2 m} + V(\hat{q}).
\eea
In the continuous space, we have that
\bea \label{eq:B2}
[\hat{q}, \hat{p}] = \mathrm{i} \hbar, \quad \hat{q} = \int dq q \ket{q}\bra{q}, \quad \hat{p} = \int dp p \ket{p}\bra{p}, \quad \bra{q} p \rangle = \frac{e^{\mathrm{i} q p / \hbar}}{\sqrt{2 \pi \hbar}},
\eea
where $\ket{q}$ and $\ket{p}$ are, respectively, the eigenvectors of $\hat{q}$ and $\hat{p}$. Note that, for the purpose of physical transparency, we retain the $\hbar$ in this section. The probability current in this case is given by the standard \cite{Messiah_v1}
\bea
\frac{d \ket{q}\bra{q}}{dt} = - \mathrm{div} J^{\mathrm{Sch}}_q, \quad J^{\mathrm{Sch}}_q = \frac{1}{2 m} \{ \hat{p}, \ket{q}\bra{q} \}.
\eea

\begin{figure}[t!]
\centering
\includegraphics[width=0.685\textwidth]{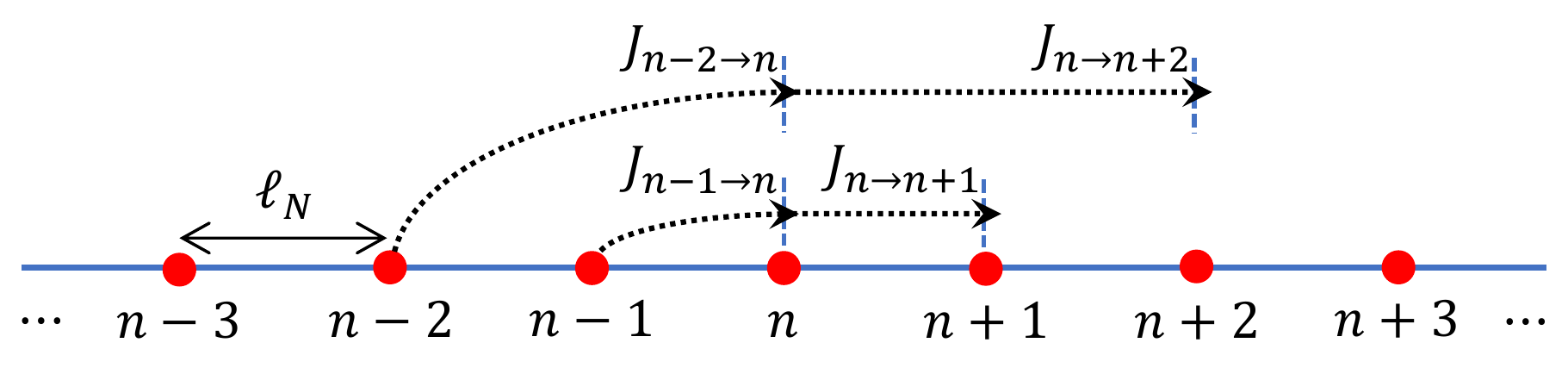}
    \caption{The discretized one-dimensional configuration space. It is illustrated how probability flows enter the site $n$ and how they leave from it.}
\label{fig:app_fig1}
\end{figure}

Let us now make the space finite and discretize it so that the allowed positions are
\bea
q_n = n \ell \epsilon_N, \quad n = -N, -N + 1, ..., N
\eea
where $\ell$ is a fixed unit of length, $\epsilon_N = \frac{1}{\sqrt{2 N + 1}}$ is an infinitesimal parameter, and $\ell_N = \ell \epsilon_N$ is the minimal length; see Fig.~\ref{fig:app_fig1}. The volume of the configuration space will then be $L_N = \ell \sqrt{2 N + 1}$, so that the $N \to \infty$ limit will guarantee both $\ell_N \to 0$ and $L_N \to \infty$. Designating the particle being at the position $q_n$ by state $\ket{q_n}$, so that $\bra{q_n} q_k \rangle = \delta_{nk}$ and $\sum_{n} \ket{q_n}\bra{q_n} = \mathbb{I}_{2 N + 1}$, where $\mathbb{I}_{2 N + 1}$ is the identity operator in the $(2 N + 1)$-dimensional Hilbert space, we can introduce the discretized position operator as
\bea
Q_N = \ell \epsilon_N \sum_n n \ket{q_n}\bra{q_n}.
\eea
Furthermore, introducing
\bea
P_N = 2 \pi \hbar \ell^{-1} \epsilon_N \sum_n n \ket{p_n}\bra{p_n},
\eea
with the eigenstates $\ket{p_n}$ determined through
\bea \label{eq:B6}
\bra{q_k} p_n \rangle = \epsilon_N e^{2 \pi \mathrm{i} \epsilon_N^2 k n},
\eea
it is straightforward to see that $\bra{p_n} p_k \rangle = \delta_{nk}$ and $\sum_n \ket{p_n}\bra{p_n} = \mathbb{I}_{2N + 1}$, and that $P_N$ serves as a translation operator for the coordinate:
\bea \label{eq:B7}
e^{- \frac{\mathrm{i} P_N \ell_N}{\hbar}} \ket{q_n} = \ket{q_{n+1}} = \ket{q_n + \ell_N}.
\eea
Analogously,
\bea
e^{\frac{\mathrm{i} Q_N \frac{2 \pi \hbar}{\ell} \epsilon_N}{\hbar}} \ket{p_n} = \ket{p_{n+1}} = \ket{p_n + 2 \pi \hbar \ell^{-1} \epsilon_N},
\eea
where $2 \pi \hbar \ell^{-1} \epsilon_N$ is the discretization step for the momentum variable.

Now, in the limit of $N \to \infty$, the minimal step, $\ell_N = \frac{\ell}{\sqrt{2 N + 1}}$, goes to zero, meaning that essentially any $q$ and $p$ correspond to some eigenvalues of operators $Q_N$ and $P_N$, and the translation operators, $e^{-\frac{\mathrm{i} q P_N}{\hbar}}$ and $e^{\frac{\mathrm{i} p Q_N}{\hbar}}$, act as continuous translations of, respectively, the eigenvectors of $Q_N$ and $P_N$. Since the above form and action of translation operators can be shown to be equivalent to the canonical commutation relations \cite{Messiah_v1}, we conclude that $N \to \infty$ is a ``physically'' valid continuous limit, and $Q_N \to \hat{q}$ and $P_N \to \hat{p}$. Note the slight discrepancy between the last relation in Eq.~\eqref{eq:B2} and Eq.~\eqref{eq:B6} stems from the fact that the normalization condition in the continuous space is $\bra{q} q \rangle = \delta(0)$ (where $\delta$ is the Dirac's $\delta$-function), whereas in the discrete case we took $\bra{q_n} q_n \rangle = 1$. If we define $\widetilde{\ket{q_n}}= (\ell_N)^{-1/2} \ket{q_n}$ and $\widetilde{\ket{p_n}}= \left(2 \pi \hbar \ell^{-1} \epsilon_N \right)^{-1/2} \ket{p_n}$, then $\widetilde{\langle q_k} | \widetilde{p_n \rangle} = \frac{1}{\sqrt{2 \pi \hbar}} e^{\mathrm{i} q_k p_n / \hbar}$ and $\widetilde{\ket{q_n}} \to \ket{q}$ and $\widetilde{\ket{p_n}} \to \ket{p}$ as $N \to \infty$.

We are now in the position to perform the action (c). In order to do that, let us first calculate $J_{n \to n + k}$ ($k\in \mathbb{N}$):
\bea
J^{(\mathrm{tun})}_{n \to n + k} &=& \frac{\mathrm{i}}{\hbar} (\ket{q_n}\bra{q_n} H_{cN} \ket{q_{n+k}}\bra{q_{n+k}} - \ket{q_{n+k}}\bra{q_{n+k}} H_{cN} \ket{q_n}\bra{q_n})
\\
&=& \frac{\mathrm{i}}{2 m \hbar}(\ket{q_n}\bra{q_n} P_N^2 \ket{q_{n+k}}\bra{q_{n+k}} - \ket{q_{n+k}}\bra{q_{n+k}} P_N^2 \ket{q_n}\bra{q_n}).
\eea
Using Eqs.~\eqref{eq:B6} and \eqref{eq:B7}, we can write
\bea
\bra{q_n} P_N^2 \ket{q_{n+k}} = \frac{4 \pi^2 \hbar^2}{\ell^2 \epsilon_N^2} \sum_{m = -N}^N \epsilon_N^2 (\epsilon_N^2 m)^2 e^{-2 \pi \mathrm{i} (\epsilon_N^2 m) k}, \eea
where we immediately see that
\bea
\sum_{m = -N}^N \epsilon_N^2 (\epsilon_N^2 m)^2 e^{-2 \pi \mathrm{i} (\epsilon_N^2 m) k} \; \overset{N \to \infty}{\longrightarrow} \; \int_{-1/2}^{1/2} dx x^2 e^{-2 \pi \mathrm{i} x k} = \frac{(-1)^k}{2 \pi^2 k^2},
\eea
leading us to
\bea
\bra{q_n} P_N^2 \ket{q_{n+k}} \overset{N \to \infty}{\longrightarrow} \frac{2 \hbar^2}{\ell_N^2} \frac{(-1)^k}{k^2},
\eea
which means that
\bea \label{eq:B14}
J^{(\mathrm{tun})}_{n \to n + k} \overset{N \to \infty}{\longrightarrow} \frac{\mathrm{i} \hbar}{m \ell_N^2} \frac{(-1)^k}{k^2} (\ket{q_n}\bra{q_{n+k}} - \ket{q_{n+k}}\bra{q_n});
\eea
note the similarity of this expression and Eq.~\eqref{eq:MODELtuncur}.

In order to perform the final action (d), we take $N \gg 1$ and note that Eq.~\eqref{eq:div3} can be rewritten as follows:
\bea
\frac{d \ket{q_n}\bra{q_n}}{d t} &=& - \sum_{k\geq 1} (J^{(\mathrm{tun})}_{n \to n + k} - J^{(\mathrm{tun})}_{n - k \to n})
\\ \label{eq:B16}
&=& - \sum_{k\geq 1} k \ell_N \frac{J^{(\mathrm{tun})}_{n \to n + k} - J^{(\mathrm{tun})}_{n - k \to n}}{k \ell_N}.
\eea
Taking into account that $J^{(\mathrm{tun})}_{n \to n + k}$ is the current entering site $n+k$ and $J^{(\mathrm{tun})}_{n - k \to n}$ is the current entering site $n$, in the limit of $\ell_N \to 0$, for any fixed $k$, the fraction in Eq.~\eqref{eq:B16} tends to the divergence of the ``$k$th-neighbour'' current $J^{(\mathrm{tun})}_{n \to n + k}$ at the position $n$. Therefore, we can write
\bea
\frac{d \ket{q_n}\bra{q_n}}{d t} \overset{N \to \infty}{\longrightarrow} - \mathrm{div} J(q_n),
\eea
where
\bea
J(q_n) = \sum_{k=1}^\infty k \ell_N J^{(\mathrm{tun})}_{n \to n + k}.
\eea
Now, noticing from Eq.~\eqref{eq:B7} that
\bea \label{eq:B15}
\ket{q_{n+k}} = e^{\frac{P_N \ell_N}{\mathrm{i} \hbar} k} \ket{q_n}
\eea
and substituting into Eq.~\eqref{eq:B14}, we obtain
\bea \label{eq:B20}
J(q_n) = \frac{\mathrm{i} \hbar}{m \ell_N} \left( \zeta \ket{q_n}\bra{q_n} - \ket{q_n}\bra{q_n} \zeta^\dagger \right),
\eea
where
\bea \label{eq:B21}
\zeta = \sum_{k = 1}^\infty \frac{(-1)^{k-1}}{k} e^{\frac{P_N \ell_N}{\mathrm{i} \hbar} k} = \ln \left( \mathbb{I}_{2N+1} + e^{\frac{P_N \ell_N}{\mathrm{i} \hbar}} \right).
\eea
Here, in obtaining the second equation, we used that $\ln(1 + x) = \sum_{k=1}^\infty \frac{(-1)^{k-1} x^k}{k}$. Furthermore, since $\epsilon_N \ll 1$, we can Taylor-expand the logarithm in Eq.~\eqref{eq:B21} with respect to $\ell_N$:
\bea
\zeta = \ln 2 \, \mathbb{I}_{2N+1} + \frac{P_N \ell_N}{2\mathrm{i} \hbar} + O(\ell_N^2),
\eea
which, substituted into Eq.~\eqref{eq:B20}, leads to
\bea
J(q_n) \overset{N \to \infty}{\longrightarrow} \frac{1}{2 m} \left\{ P_N, \ket{q_n}\bra{q_n} \right\}.
\eea

Finally, transitioning from $\ket{q_n}$ to $\widetilde{\ket{q_n}}$, we have
\bea
\frac{d \widetilde{\ket{q_n}} \widetilde{\bra{q_n}}}{d t} = - \mathrm{div} \widetilde{J}(q_n),
\\
\widetilde{J}(q_n) \overset{N \to \infty}{\longrightarrow} \frac{1}{2 m} \left\{ P_N, \widetilde{\ket{q_n}} \widetilde{\bra{q_n}} \right\} \overset{N \to \infty}{\longrightarrow} \frac{1}{2 m} \{ \hat{p}, \ket{q}\bra{q} \}.
\eea
We have thus proved that our discrete-space theory has a proper continuous limit and, in that limit, for closed systems, reproduces the well-known expression for the quantum probability current.

Analogously to the tunneling current, by performing the (a)-(b)-(c)-(d) procedure and using the techniques developed in this section, our discrete-space formalism can be extended to general open continuous-variable systems. For example, one can derive the continuous limit of the environment-induced current given by Eq.~\eqref{eq:thcur} for open continuous-variable systems evolving according to GKLS MEs; this however is an analysis that will be presented elsewhere.

\section{Current of a Markovian particle on a triangle}
\label{app:triangle}

The graph is a triangle, therefore the continuity equation reduces to
\bea
\frac{d x_j}{d t} = J_{j-1 \to j} - J_{j \to j+1}.
\eea
Moreover, we can write $x_{j+3} = x_j$, and, therefore, for any $j$, we have that
\bea \label{eq:idres}
x_{j-1} + x_{j} + x_{j+1} = \mathbb{I}.
\eea
Now, the time derivative of $x_j$ is given by $\frac{d x_j}{d t} = \mathrm{i}[H, x_j] + \sum_\lambda \gamma_\lambda \mathcal{S}^{*}[\Lambda_\lambda][x_j]$ , where $S^{*}$ is the dual of $S$: $S^{*}[\Lambda][x] = \Lambda^\dagger x \Lambda - \frac{1}{2} \{ \Lambda^\dagger \Lambda, x \}$. Using the above identity resolution in $[H, x_j] = \mathbb{I} H x_j - x_j H \mathbb{I}$, we immediately obtain that
\bea
\mathrm{i}[H, x_j] = J^{(\mathrm{tun})}_{j-1 \to j} - J^{(\mathrm{tun})}_{j \to j+1},
\eea
where $J^{(\mathrm{tun})}_{j-1 \to j}$ is given by Eq.~\eqref{eq:tuncur}.

Turning to the dissipative part of $\frac{d x_j}{d t}$, $\mathcal{D}^*[x_{j}] = \sum_{\lambda} \gamma_\lambda \Big[ \Lambda_\lambda^\dagger x_{j} \Lambda_\lambda - \frac{1}{2} \! \left\{ \! \Lambda_\lambda^\dagger \Lambda_\lambda, x_{j} \! \right\} \! \Big]$, let us substitute the identity resolution (Eq.~\eqref{eq:idres}) in the trivial identities $\Lambda_\lambda = \Lambda_\lambda \mathbb{I}$ and $\Lambda_\lambda^\dagger = \mathbb{I} \Lambda_\lambda^\dagger \mathbb{I}$, in order to represent $\mathcal{D}^*[x_{j_a}]$ as a sum of terms of the form $x_{j'} \Lambda_\lambda^\dagger x_{j''} \Lambda_\lambda x_{j'''}$. As can be checked by direct inspection, all these terms can be rearranged in such a way that
\bea \label{eq:div2}
\mathcal{D}^*[x_{j}] = J^\mathrm{(th)}_{j - 1 \to j} - J^\mathrm{(th)}_{j \to j + 1},
\eea
with
\bea \label{eq:thercur1}
J^\mathrm{(th)}_{j \to j + 1} =  \sum_{\lambda} \gamma_\lambda && \Big[ (j, j + 1, j) - (j + 1, j, j + 1)
\\   \label{eq:thercur2}
&& + \; \frac{1}{2} (j + 1, j + 1, j) - \frac{1}{2}(j, j, j + 1)
\\   \label{eq:thercur3}
&& + \; \frac{1}{2} (j, j + 1, j + 1) - \frac{1}{2}(j + 1, j, j)
\\   \label{eq:thercur4}
&& + \; \frac{1}{2} (j, j + 1, j + 2) - \frac{1}{2}(j + 1, j, j + 2)
\\   \label{eq:thercur5}
&& + \; \frac{1}{2} (j + 2, j + 1, j) - \frac{1}{2}(j + 2, j, j + 1)  \Big],
\eea
where the symbol $(j', j'', j''')$ is introduced for ease of notation and stands for $x_{j'} \Lambda_\lambda^\dagger x_{j''} \Lambda_\lambda x_{j'''}$. Using the identity resolution \eqref{eq:idres} again, we can further massage the above expression into
\bea \label{eq:thercur}
J^\mathrm{(th)}_{j \to j + 1} =  \frac{1}{2} \sum_{\lambda} \gamma_\lambda \left[ \{ x_j,  \Lambda_\lambda^\dagger x_{j + 1} \Lambda_\lambda \} - \{ x_{j+1}, \Lambda_\lambda^\dagger x_j \Lambda_\lambda \} \right], 
\eea
which exactly coincides with Eq.~\eqref{eq:thcur}.

This calculation thus demonstrates that no terms, other than those in Eq.~\eqref{eq:current}, arise when one derives the current of a particle on a triangle, directly stemming from the equations of motion.

\section{Two-strong-measurement current}
\label{app:TSM}

Let us see what the current of a particle described by a GKLS master equation will look like if we substitute the first weak measurement (see Eq.~\eqref{eq:weakval} and the discussion around it) by a standard (i.e., non-weak) quantum von Neumann measurement \cite{mikeike, bp}. To find the current, we first measure the particle at vertex $j$, at the moment of time $t$. The probability of finding the particle at the measured vertex will be given by $p_{j}(t)=\tr(x_{j} \rho(t))$, and, immediately after the measurement, the state of the system will collapse into $\rho_{j}=\frac{x_{j} \rho(t) x_{j}}{p_{j}}$. Within the period of time $\epsilon$, the state will evolve into $\rho_{j} +\epsilon \mathcal{L}[\rho_{j}] + O(\epsilon^2)$, where $L$ is the GKLS generator. Therefore, the probability of finding the particle at site $x_{j'}$, at the moment of time $t + \epsilon$, will be $\tr \big[ x_{j'} \big( \rho_{j} + \epsilon \mathcal{L}[\rho_{j}] + O(\epsilon^2)\big) \big]$. Taking the reverse flow into account, we thus find the two-strong-measurement current to be given by
\bea  \nonumber
\langle J^{(\mathrm{TSM})}_{j\to j'}\rangle && = \tr \big( \mathcal{L}[x_{j} \rho x_{j}] x_{j'} \big) - (j \leftrightarrow j')
\\ \label{eq:TSMcur}
&& = \sum_\lambda \gamma_\lambda [\tr(\Lambda_\lambda x_j \rho x_j \Lambda_\lambda^\dagger x_{j'}) - (j \leftrightarrow j')].
\eea
For the particle on a triangle, we immediately see that $\langle J^{(\mathrm{TSM})}_{i\to i'}\rangle$ coincides with the RHS of Eq.~\eqref{eq:thercur1}. In other words, making the first measurement strong eliminates all coherent contributions to the current, which are given by Eq.~\eqref{eq:tuncur} and Eqs.~\eqref{eq:thercur2}-\eqref{eq:thercur5}. We stress that Eq.~\eqref{eq:tuncur} is the tunneling current.

\section{Symmetries and current in classical and local quantum MEs}
\label{app:symm}

Let us start by analysing the steady-state regime in the classical setting. There, the evolution is given by
\bea \label{eq:clasme}
\frac{\partial p_{j_a j_b}}{\partial t} = && \sum_{j_a' = j_a \pm 1} \big[ p_{j_a' j_b} W_{j_a' j_b \vert j_a j_b} - p_{j_a j_b} W_{j_a j_b \vert j_a' j_b} \big]
\\
&& + \sum_{j_b' = j_b \pm 1} \! \big[ p_{j_a j_b'} W_{j_a j_b' \vert j_a j_b} - p_{j_a j_b} W_{j_a j_b \vert j_a j_b'} \big],
\eea
where $W_{j_a,j_b \vert j_a',j_b'}$ are the transition rates satisfying local detailed balance conditions:
\bea \label{eq:trrate1}
\frac{W_{j_a j_b \vert j_a' j_b}}{W_{j_a' j_b \vert j_a j_b}} = e^{- \beta_a \left[ U_{j_a' j_b} - U_{j_a j_b} \right]}, \quad \; \frac{W_{j_a j_b \vert j_a j_b'}}{W_{j_a j_b' \vert j_a j_b}} = e^{- \beta_b \left[ U_{j_a j_b'} - U_{j_a j_b} \right]} .
\eea
Upon collecting $p_{j_a,j_b}$'s and the transition rates into, respectively, the vector $p$ and matrix $\mathcal{W}$, Eq.~\eqref{eq:clasme} can be rewritten as $\frac{dp}{dt} = \mathcal{W} p$. 

The stochastic matrix $\mathcal{W}$ depends on the potential $U$ through Eq.~\eqref{eq:trrate1} (these can be chosen to be given by, e.g., Eq.~\eqref{eq:trrate3}).

Now, since the phase space is discrete, all transformations are given by permutation matrices. A configuration function, e.g., the steady state $p$, is symmetric under a transformation $P$ if $Pp=p$. Importantly, Eqs.~\eqref{eq:clasme}-\eqref{eq:trrate1} ensure that $U$ and $p$ have the same symmetries. Note, however, that, although $\mathcal{W}$ and $U$ might have common symmetry transformations, due to the nonlinear relation between $\mathcal{W}$ and $U$, not all symmetries of $U$ will in general be respected by $\mathcal{W}$. The global ``rigid'' rotation symmetry of the potential (namely, $U_{i+k,j+k} = U_{i, j}$) provides an example of the first kind: $\mathcal{W}$ is also symmetric under the global rotation, which is expressed in the fact that $\mathcal{W}_{j_a+1 \, j_b+1, j_a'+1 \, j_b'+1} = \mathcal{W}_{j_a j_b, j_a' j_b'}$. Interestingly, $\mathcal{W}$ maintains a single steady state despite this symmetry. As is discussed below, this is related to the fact that, although the local GKLS equation giving rise to the classical evolution is, as a whole, symmetric under global rotations, its individual jump operators are not. Following the terminology of Ref.~\cite{Buca_2012}, one may say that the classical dynamics is hence only ``weakly'' symmetric under global rotations.

\begin{figure}[t!]
\centering
\includegraphics[width=0.65\textwidth]{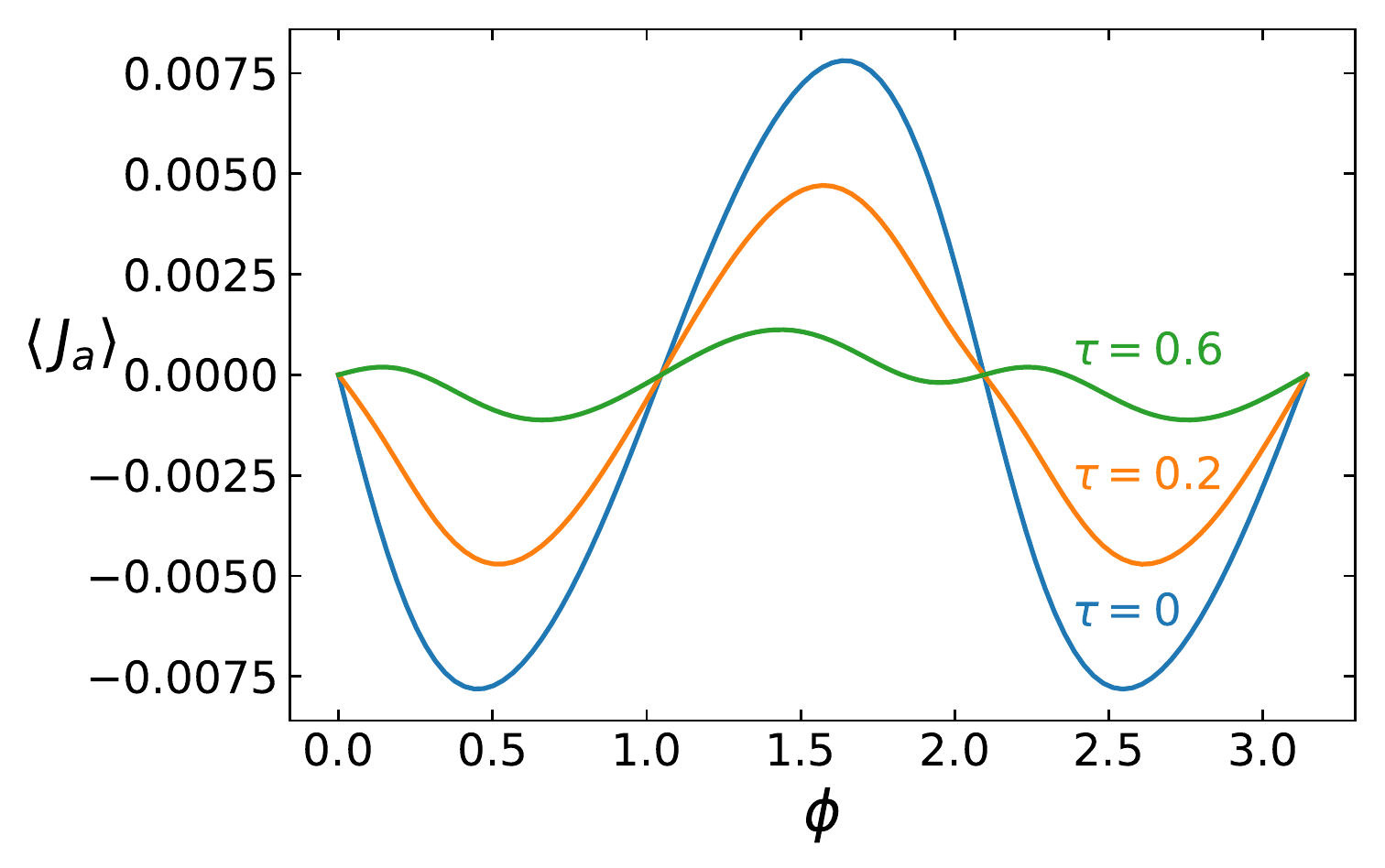}
    \caption{The average current of particle $a$, $\langle J_a\rangle$, versus the phase $\phi$ in the steady state of the local GKLS ME. The blue curve corresponds to zero tunneling and is equivalent to the classical case. The orange and green curves correspond to, respectively, $\tau=0.2$ and $\tau=0.6$. The other parameters are: $T_a = 0.2$, $T_b = 1$, $\gamma_a = \gamma_b = 0.2$, $K = 2$. On the side $a$, where the temperature is lower, the quantum effects of the tunneling become more dominant, and, with the increase of the tunneling rate, are expressed by changes in the current's direction. No similar inversions in the direction of the current are observed for particle $b$, which is attached to a hotter-temperature bath. As it can be clearly seen, the current is zero whenever $\phi=k\pi/3$.}
\label{fig:app_fig2}
\end{figure}

Now, it can be shown numerically \cite{Fogedby_2017} that there is no current in the system whenever $\phi = k \pi/3$ ($k$ is an arbitrary integer), and the current is non-zero whenever $\phi \neq k \pi/3$ (see Fig.~\ref{fig:app_fig2}). At the same time, we notice that, whenever $\phi = k \pi/3$, $U_{j,i} = U_{i, j + k}$. In other words, when $\phi = k \pi/3$, the potential is symmetric under the combination of particle exchange and a unilateral rotation. The disappearance of the current in this case can be explained by a simple physical argument. Indeed, exchanging the particles is equivalent to swapping the temperatures of the baths. On the other hand, taking, for simplicity, $\phi=0$, we see that the system and its state remain intact (recall that the steady state inherits all $U$'s symmetries). Put otherwise, inverting the direction of the temperature gradient does not alter the currents. Now, drawing our intuition from phenomenological non-equilibrium thermodynamics \cite{deGrootMazur}, where, for small temperature differences, the current is a linear function of the temperature gradient, we thus conclude that the current must be zero.

Interestingly, the generator of the evolution, $\mathcal{W}$, is not symmetric under the above ``generalized'' exchange symmetry. Nevertheless, the increased degeneracy in $U$'s spectrum ($U$ has $2$ distinct eigenvalues when $\phi=k\pi/3$, and $3$ when $\phi\neq k\pi/3$) brings about degeneracies in $\mathcal{W}$, leaving it with only $5$ distinct eigenvalues (contrasting $\mathcal{W}$'s having no degeneracies when $\phi\neq k\pi/3$), which in turn means that $\mathcal{W}$ is more symmetric when $\phi=k\pi/3$.

The classical ME and phenomenological quantum ME, 
\bea \label{eq:GKLSloc}
\frac{\partial\rho}{\partial t} = \mathcal{L}_\mathrm{loc}[\rho] = - \mathrm{i}[H,\rho] && + \sum_{j_a,j_a',j_b} W_{j_a j_b \vert j_a' j_b} \mathcal{S}[L_{j_a,j_a',j_b}][\rho]
\\
&& + \sum_{j_a,j_b,j_b'} W_{j_a j_b \vert j_a j_b'} \mathcal{S}[L_{j_a,j_b,j_b'}][\rho],
\eea
where $L_{j_a, j_a', j_b} = \ket{j_a, j_b} \bra{j_a', j_b}$, $L_{j_a,j_b,j_b'} = \ket{j_a, j_b} \bra{j_a, j_b'}$, share many similarities. Indeed, the phenomenological ME is only weakly symmetric under global rotation $R$: It is easy to see that $R$ does not commute with the jump operators $L_\mu$, whereas it can be seen by direct inspection that $R\mathcal{L}_\mathrm{loc}[\rho] R^\dagger = \mathcal{L}_\mathrm{loc}[R \rho R^\dagger]$. A GKLS ME, only weakly symmetric under a non-trivial unitary transformation, does not necessarily have multiple steady states \cite{Buca_2012}, and our phenomenological ME, given by Eq.~\eqref{eq:GKLSloc}, indeed has a unique steady state for any value of $\tau$. This can be seen as an artifact of the global (system plus baths) Hamiltonian not being symmetric under global rotation.

As we can see in Fig.~\ref{fig:app_fig2}, where the average of $J_{j_a\to j_a+1}$ in the steady state (the same for all $j_a$'s) is plotted against $\phi$, the current is again zero whenever $\phi=k\pi/3$. The generalized exchange symmetry of the potential, discussed in the previous section and, in the quantum case, given by $\Xi_k = \mathrm{SWAP} \cdot \left( \mathbb{I} \otimes R_b^k \right)$, is also a symmetry of $H_{ab}$: $[H, \Xi_k]=0$. Similarly to the classical case, although the generator of the evolution is not symmetric under $\Xi_k$ ($\Xi_k \mathcal{L}_\mathrm{loc}[\rho] \Xi_k^\dagger \neq \mathcal{L}_\mathrm{loc}[\Xi_k \rho \Xi_k^\dagger]$), the steady state is. Moreover, taking into account that, in the steady state, unilateral rotation does not affect the average local current ($J_{j_b \to j_b'} = J_{j_b + 1 \to j_b' + 1}$), we conclude that the transformation $\Xi_k$ is equivalent to swapping the temperatures of the baths. Hence, by the same physical argument as in the classical case, we restore the intuition as to why the current is zero for $\phi=k\pi/3$. Note that the situation is slightly different for global GKLS ME (see, e.g., Fig.~\ref{fig:fig2}): only the thermal current nullifies at $\phi = k \pi/3$, whereas the tunneling current can be non-zero. This shows that, when all quantum effects are properly taken into account, the rigid link between symmetry and current is no longer present, and the link is broken by the purely quantum component -- the tunneling current.

Importantly, we note that the average steady-state current is $2\pi/3$-periodic. Indeed, the potential satisfies $U_{j_a, j_b}(\phi + 2\pi/3) = U_{j_a, j_b + 1}(\phi)$, which, along with the fact that the tunneling parts of the Hamiltonian are local-rotation-invariant, means that a $2\pi/3$ phase shift is equivalent to a unilateral rotation. On the other hand, the average local current in the steady state is invariant under local rotation, which proves the $2\pi/3$-periodicity of the average current. This periodicity is evident in Fig.~\ref{fig:app_fig2}.

Interestingly, as can be read off from Fig.~\ref{fig:app_fig2}, the current inversion phenomenon observed for the global ME in the main text, takes place also in this case. This is a purely quantum phenomenon, and it occurs only when $T_a$ is sufficiently low.

As a final remark, let us note that the thermodynamic inconsistency of the local GKLS manifests itself when $T_a=T_b$. Indeed, when $\tau \neq 0$, the steady-state solution, $\rho^{\mathrm{loc}}_\mathrm{st}$, is not a thermal state at temperature $T_a$; in fact, $\left\Vert \rho^{\mathrm{loc}}_\mathrm{st} - \frac{1}{Z} e^{-H_{ab}/T_a} \right\Vert\propto \tau$ (where $Z=\tr e^{-H/T_a}$), so the deviation cannot be considered to be small. Interestingly, however, this inconsistency affects the current only mildly: $\langle J_a\rangle \propto \tau^2$, which is only a second-order effect, and, moreover, $\langle J_a \rangle + \langle J_b \rangle = 0$.

\section{Contextuality explained}
\label{app:context}

Here, we will briefly outline the main concepts behind the notion of contextuality. See Refs.~\cite{Spekkens_2005, Pusey_2014, Lostaglio_2018} for an extensive discussion. An operational theory (e.g., quantum mechanics) consists of two processes: preparations ($\Pi$) and measurements ($M$). Given a preparation (which, in quantum mechanics, is fully characterized by a density matrix), the theory predicts the probability distribution of the outcomes ($o$) of measurements: $\mathcal{P}(o|M, \Pi)$. A classical stochastic system is said to be a hidden-variable (aka ontological) model for an operational theory, if, to any system, it associates a space of real (or ontic) states, $\{ \zeta \}$, such that each state in that space fully characterizes the system. To any preparation $\Pi$ of the system, it associates a probability distribution over the state space, $\nu_\Pi(\zeta)$ (called epistemic state). And since the ontic states fully characterize the system, for each ontic state, there exists a probability distribution of outcomes of a measurement: $\pi_{o|M}(\zeta)$ (called response function). Finally, the hidden variable model should reproduce the statistics of measurements of the system:
\bea
\mathcal{P}(o|M, \Pi) = \int d \zeta \pi_{o|M}(\zeta) \nu_\Pi(\zeta).
\eea

Furthermore, a hidden-variable model is measurement noncontextual, if to operationally equivalent measurements, $M' \sim M$, it prescribes identical response functions: $\pi_{o|M'}(\zeta) = \pi_{o|M}(\zeta)$. Analogously, a hidden-variable model is preparation noncontextual, if to operationally equivalent preparations, $\Pi' \sim \Pi$, it prescribes identical epistemic states: $\nu_{\Pi'}(\zeta) = \nu_\Pi(\zeta)$.

Now, if the operational theory is capable of producing a probability distribution $\mathcal{P}$ such that no measurement (preparation) noncontextual hidden variable model can reproduce it, then that situation is said to be measurement (preparation) contextual.

Taking a slightly less formal stance, in the effort to separate quantum from classical, let us assume that the ontic model is represented by classical mechanics. Since in classical mechanics full characterization excludes probabilistic description, then response functions should be either $0$ or $1$. Hidden-variable models with such response functions are called outcome deterministic. In short, we expect classical hidden-variable models to be measurement and preparation noncontextual and outcome deterministic.

The following theorem (first proven for pure states in Ref.~\cite{Pusey_2014}, then generalized to mixed state in Ref.~\cite{Lostaglio_2018}) holds: If $\rho$ is a quantum state and $x$ and $x'$ are projectors, such that $\tr(\rho x x')<0$, then there exists a POVM, $\{ V_\mu \}$, such that no measurement-noncontextual outcome-deterministic hidden-variable model can simulate the preparation $\rho$, the measurement $\{ V_\mu \}$, and postselection $x_{j'}$. See Refs.~\cite{Pusey_2014, Lostaglio_2018} for the proof.

\section{$2\pi/3$-periodicity of current, heat, and entanglement}
\label{app:period}

Let us prove that the steady-state current, heat flux, and negativity are all $2\pi/3$-periodic function of $\phi$. All these stem from the fact that, since the potential satisfies $U_{j_a, j_b}(\phi + 2\pi/3) = U_{j_a, j_b + 1}(\phi)$ and the tunneling parts of the rotor Hamiltonian are invariant under local rotation,
\bea \label{eq:cocodrilo}
H_{ab}(\phi + 2 \pi /3) = I\otimes R_b^\dagger \, H_{ab} \, I \otimes R_b.
\eea
Furthermore, Eq.~\eqref{eq:cocodrilo} and $[X, R_b]=0$ mean that $\phi\to\phi+2\pi/3$ leads to $\Lambda_\alpha(\omega) \to \mathbb{I}_a\otimes R_b^\dagger \Lambda_\alpha(\omega) \mathbb{I}\otimes R_b$ (see Eqs.~\eqref{eq:jumpglob} and \eqref{eq:As}). Hence, it also holds that
\bea \label{eq:cocodrilo2}
\rho^\mathrm{st}(\phi + 2\pi/3) = \mathbb{I}_a\otimes R_b^\dagger \rho^\mathrm{st} \mathbb{I}\otimes R_b.
\eea
In view of the fact that the average local current in the steady state is invariant under local rotation, this proves the $2\pi/3$-periodicity of the average current.

For the heat flux, let us notice that Eq.~\eqref{eq:cocodrilo2} implies that $\phi \to \phi + 2\pi/3$ leads to $\mathcal{D}_\alpha[\rho^\mathrm{st}] \to \mathbb{I}_a\otimes R_b^\dagger \mathcal{D}_\alpha[\rho^\mathrm{st}] \mathbb{I}\otimes R_b$, which, through Eq.~\eqref{eq:Qdot}, proves that $\phi\to\phi+2\pi/3$ indeed leaves $\dot{Q}$ unchanged.

Lastly, in view of the fact that the negativity is invariant under local unitary transformations \cite{4H_2009}, Eq.~\eqref{eq:cocodrilo2} immediately proves that the negativity is indeed $2\pi/3$-periodic.

\section{Trading local coherence for entanglement}
\label{app:titfortat}

\begin{figure}[t!]
\centering
\includegraphics[width=0.65\textwidth]{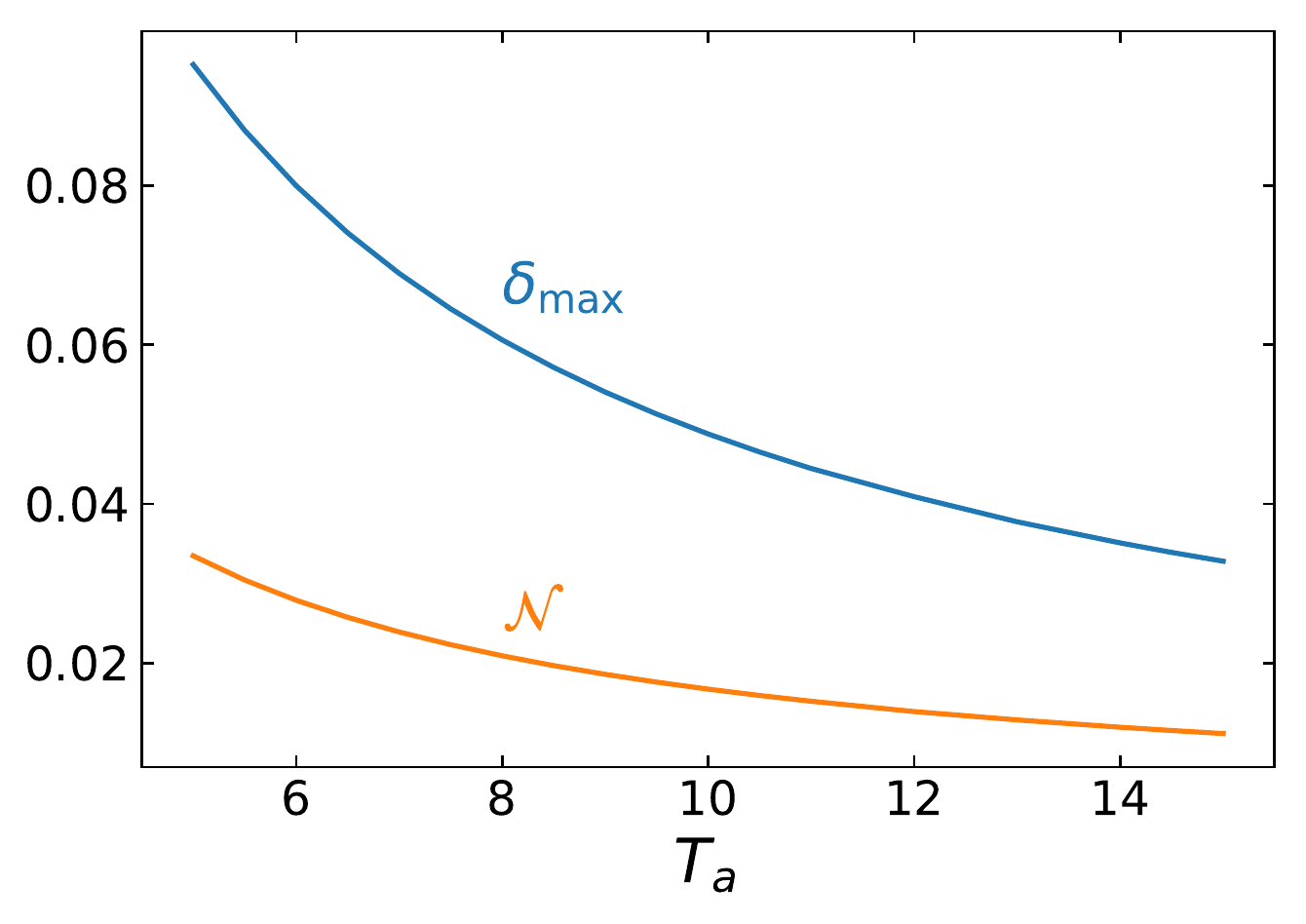}
    \caption{Entanglement and initial purity. The minimal purity (as quantified by $\delta_{\max}$) and the negativity $\mathcal{N}$ of the steady state for $\delta = \delta_{\max}/2$, plotted against $T_a$ ($T_b = 2T_a$). The tunneling rate $\tau$ is chosen to be $0.1$. For this value of $\tau$, $T[H] \approx 0.3948$, which is much smaller than the bath temperatures in the plot. The phase $\phi$ is set to the value $\pi/3$ and all the other parameters are the same as those in Fig.~\ref{fig:app_fig2}.}
\label{fig:app_fig3}
\end{figure}

As is discussed around Eq.~\eqref{eq:Barbados}, for arbitrary initial state $\rho_0$, $\rho^\mathrm{st} = \sum_{k=1}^3 \lambda_k \rho^\mathrm{st}_k$, with $\lambda_{1,2} = \frac{1}{3} - \frac{2}{3} \mathrm{Re}(e^{\pm \mathrm{i} \pi / 3} \theta_0)$, $\lambda_3=1-\lambda_1 - \lambda_2$, where $\theta_0=\sum_{j_a j_b}\bra{j_a,j_b}\rho_0 \ket{j_a+1, j_b+1}$. For uncorrelated initial states, i.e., when $\rho_0 = \sigma_a \otimes \sigma_b$, we have that $\theta_0 = \theta_{\sigma_a} \theta_{\sigma_b}$, where $\theta_\sigma = \sigma_{12} + \sigma^*_{13} + \sigma_{23}$. When mixing the basis steady states, the entanglement vanishes, especially when the states are only weakly entangled. Therefore, in the limit of high $T_a$ and $T_b$, in order for $\rho^\mathrm{st}$ to be entangled, one of the $\lambda_k$'s has to be close to $1$. For definiteness, let us pick $\lambda_1$; the analysis for $\lambda_2$ and $\lambda_3$ will be identical. Then, requiring $\lambda_1 \approx 1$ will be equivalent to
\bea \label{eq:Hawaii}
|\theta_\sigma||\theta_\kappa| \cos(\arg\theta_\sigma + \arg\theta_\kappa + 4\pi/3) \approx 1.
\eea
On the other hand, for any density matrix $\sigma$ in a $3$-dimensional Hilbert space, we have that $|\theta_\sigma| \leq \sum_{j < j'}|\sigma_{jj'}| \leq \sqrt{3}\sqrt{\sum_{j<j'} |\sigma_{jj'}|^2}=\sqrt{\frac{3}{2}\left[ \tr(\sigma^2) - \sum_j \sigma_{jj}^2 \right]}$, where, in the second step, we used the Cauchy-Schwarz inequality. Using the latter again, it is easy to see that, since $\sigma_{jj}\geq 0$ and $\sum_j \sigma_{jj}=1$, $\sum_j \sigma_{jj}^2\geq 1/3$, which leaves us with $|\theta_\sigma| \leq \sqrt{\frac{3}{2}\tr(\sigma^2) - \frac{1}{2}} \leq 1$. We thus see that the quantity $\tr(\sigma^2)$, also known as the purity of the state \cite{mikeike}, controls the upper bound on $|\theta_\sigma|$. For Eq.~\eqref{eq:Hawaii}, $\theta_\sigma \leq 1$ means that all multipliers must be $\approx 1$, which means that, in order for Eq.~\eqref{eq:Hawaii} to hold, both $\sigma_a$ and $\sigma_b$ must be nearly pure (the higher the $T_a$, the purer). The states $\ket{\sigma_a} = \frac{1}{\sqrt{3}} \left( \ket{1} + e^{-2\pi\mathrm{i}/3} \ket{2} + e^{2\pi\mathrm{i}/3} \ket{3} \right)$ and $\ket{\sigma_b} = \frac{1}{\sqrt{3}}\left( \ket{1} + \ket{2} + \ket{3} \right)$ provide an example saturating all the above upper bounds. Let us now add noise to the states, e.g., via $\sigma(\delta) = (1-\delta)\ket{\sigma}\bra{\sigma} + \frac{\delta}{3} \mathbb{I}$ (the purity of this state is equal to $1 - (4\delta - \delta^2)/3$), so that they become slightly mixed. As we have just proved, there always exists an $\delta > 0$ ($\delta \to 0$ as $T_a \to \infty$) such that $\sigma_a (\delta) \otimes \sigma_b (\delta)$ evolves into an entangled steady state. For a given pair of bath temperatures, $T_a$ and $T_b$, let $\delta_{\max} (T_a, T_b)$ be the supremum of all $\delta$'s for which the steady state is entangled. In Fig.~\ref{fig:app_fig3}, we plot $\delta_{\max} (T_a, 2 T_a)$ and the negativity of the steady state for $\delta = \delta_{\max} (T_a, 2 T_a)/2$ against $T_a$. The tunneling rate for Fig.~\ref{fig:app_fig3} is chosen to be $\tau = 0.1$, and note that, for this value of $\tau$, $T[H_{ab}]\approx 0.3948$. Lastly, let us note that $\theta_\sigma$ is related to the measure of coherence in the state $\sigma$, defined as $\mathcal{C}[\sigma] = \sum_{j\neq j'} |\sigma_{jj'}|$ \cite{Baumgratz_2014}, via $\theta_\sigma \leq \frac{1}{2} \mathcal{C}[\sigma]$, and, when $\theta_\sigma$ tends to its maximum ($=1$), $\mathcal{C}[\sigma]$ is also maximized.

\section*{References}

\bibliographystyle{iopart-Karen}

\bibliography{references}

\end{document}